\pdfoutput=1 

\documentclass[twocolumn,pra,showpacs,preprintnumbers]{revtex4-1}
\usepackage{amssymb}
\usepackage{amsmath}
\usepackage{graphicx}
\usepackage{epsfig}
\usepackage{subfigure}
\usepackage{amsfonts}
\usepackage{CJK}
\usepackage{multirow}
\usepackage{amsfonts}
\usepackage{dsfont}
\usepackage{amsmath}
\usepackage{graphicx}
\usepackage{float}
\usepackage{amsmath,epsfig,amssymb,subfigure,bm,dsfont}
\usepackage{lipsum}
\usepackage[english]{babel}

\begin{document}

\title{Coordinately Assisted Distillation of Quantum Coherence in
Multipartite System}
\author{Huang-Qiu-Chen Wang}
\author{Qi Luo }
\author{Qi-Ping Su}
\author{Yong-Nan Sun}
\author{Nengji Zhou}
\author{Li Yu}
\author{Zhe Sun}
\email{sunzhe@hznu.edu.cn}
\affiliation{School of Physics, Hangzhou Normal University, Hangzhou 310036,
China}
\begin{abstract}
We investigate the issue of assisted coherence distillation in the
asymptotic limit (considering infinite copies of the resource states), by
coordinately performing the identical local operations on the auxiliary
systems of each copy. When we further restrict the coordinate operations to
projective measurements, the distillation process is branched into many sub-processes. Finally, a simple formula is given that the assisted
distillable coherence should be the maximal average coherence of the
residual states. The formula makes the experimental research of assisted
coherence distillation possible and convenient, especially for the case that
the system and its auxiliary are in mixed states. By using the formula,\ we
for the first time study the assisted coherence distillation in multipartite
systems. Monogamy-like inequalities are given to constrain the distribution
of the assisted distillable coherence in the subsystems. Taking three-qubit
system for example, we experimentally prepare two types of tripartite
correlated states, i.e., the $W$-type and GHZ-type states in a linear
optical setup, and experimentally explore the assisted coherence
distillation. Theoretical and experimental results agree well to verify the
distribution inequalities given by us. Three measures of multipartite
quantum correlation are also considered. The close relationship between the
assisted coherence distillation and the genuine multipartite correlation is
revealed.
\end{abstract}

\maketitle
\date{\today }

\section{Introduction}

Quantum coherence, as the fundamental feature of quantum mechanics and a
kind of resource\thinspace \cite{rev.mod1,rev.mod2}, is widely used in
quantum information processing\thinspace \cite{qinf}, quantum computation,
quantum algorithm\thinspace \cite{qcom1,qcom2}, quantum metrology\thinspace
\cite{metro1,metro2,metro3,metro4}, and quantum thermodynamics\thinspace
\cite{qthe1,qthe2}. It is the main reason why the quantum world is different
from the classical world\thinspace \cite{natrue.entangle}.

In order to quantify coherence\thinspace \cite{quantify}, one needs a set of
reference basis $\{|i\rangle ,\,i=0,1,2\cdot \cdot \cdot \}$, based on
which, the class of incoherent state $\mathcal{I}$ is defined with diagonal
density matrices, i.e., $\sum_{i}\rho _{i}|i\rangle \langle i|\,\in
\mathcal{I}$. Following this, incoherent operations (IO) act unchangeably on
the assemblage of all incoherent states and satisfy the map $\Lambda _{\text{%
IO}}(\mathcal{I})\subseteq \mathcal{I}$. Different types of incoherent
operations are proposed in\thinspace \cite{quantify,io1,io2,io3,io4}. A
common measure of coherence for a state $\rho $ is defined by the relative
entropy\thinspace \cite{quantify}, $C_{r}(\rho ):=\underset{\sigma \in
\mathcal{I}}{\min }S(\rho \Vert \sigma )$, to characterize the minimal
distance of $\rho $ to the class of incoherent states $\mathcal{I}$. One of
the most\ operational measure of the coherence is the distillable coherence
which is similar to the framework of the distillable entanglement\thinspace
\cite{entang.distill1,entang.distill2}, and was introduced in\thinspace \cite%
{wintedistill} at the asymptotic limit by considering infinite copies of the
state. The optimal rate of a state $\rho $ in a coherence distillation
process, defined as the distillable coherence $C_{d}(\rho )$, is evaluated
analytically $C_{d}(\rho )=C_{r}(\rho )\,$\thinspace \cite%
{wintedistill,Cr-uni view}. However, in experiments, it is a huge challenge
to collectively manipulate a large number of state copies, and to achieve
the asymptotic limit. Therefore, a kind of one-shot coherence distillation
was proposed\thinspace \cite{oneshotdistill}, and which provided the possibility
of the experimental study. This one-shot scenario was experimentally
demonstrated based on a linear optical system\thinspace \cite%
{exponeshotdistill}, where a kind of $N$-dimesional ($N\geq 2$) incoherent
operations were realized.

On the other hand, the asymptotic scenario of coherence distillation was
developed into bipartite system $\rho _{_{AB}}$, where only the operations
performed on the second party ($B$) are restricted to incoherent operations,
and the classical communication is allowed between the two parties. These
sets of operations are called local quantum-incoherent operations and
classical communications\thinspace (LQICC). Following the LQICC, the concept
of the assisted coherence distillation was established in asymptotic
settings\thinspace \cite{aoc1}. The assisted distillation rate $R$ of
subsystem $\rho _{_{B}}$ is bounded by quantum-incoherence relative
entropy\thinspace (QI relative entropy) $C_{r}^{A|B}(\rho _{_{AB}})$. For
pure states $\left\vert \Psi \right\rangle _{\text{AB}}$, the upper bound is
accessible, while for mixed states $\rho _{_{\text{AB}}}$, it is still an
open question whether the upper bound can be achieved. The experimental
study of the assisted coherence distillation was reported in\thinspace \cite%
{opticalaoc}, where the authors employed an one-copy scenario, as a way of
understanding, to experimentally simulate the case of asymptotic limit. To
overcome the difficulty of the experimental demonstration, the nonasymptotic
settings of the assisted coherence distillation was proposed\thinspace \cite%
{oneshotaoc1,oneshotaoc2}. Different from the distillation framework above,
in\thinspace \cite{fanhen} the authors introduced a scenario of steering-induced
coherence, which is defined on the eigenvectors of the considered system, and has been conveniently used in open
systems\thinspace \cite{norawsteering}.

Quantum coherence in multipartite systems has been attracted much attention
in the last decade. The problems of the quantum coherence distribution among
the constituent subsystems were considered in\thinspace \cite{mulcoh1}, and
the conversion between quantum coherence and quantum correlation\thinspace
was studied in\ \cite{mulcoh2,mulcoh3}. Following the research line of
quantum coherence, we find there is an important issue worthy of study,
i.e., how to efficiently distill coherence in a multipartite system by
choosing one of the subsystems as the assistant system.

On the other hand, we find a simple formula of the assisted coherence
distillation which is rooted in the asymptotic framework of the coherence
distillation and is suitable for mixed states. We employ a class of
operations named as coordinate local quantum-incoherent operations and
classical communication\thinspace (CoLQICC), which is a subset of LQICC.
With this type of operations, we define the coordinately assisted
distillation of coherence and obtain a simple analytical formula, which
overcomes the difficulty of infinite copy limit and facilitates the
experimental study. By using the formula, we for the first time investigate
the distribution of the assisted distillable coherence in multipartite
systems and develop a monogamy-like inequality to show that the assisted
distillable\ coherence of individual subsystems should be constrained by
that of the whole remaining part. Taking a three-qubit system for example, we
experimentally demonstrate the distribution of the assisted coherence distillation in a linear optical setup.


\section{Coordinately Assisted Distillation of Coherence}

For a bipartite system of Alice and Bob sharing the state $\rho _{_{AB}}$,
the aim of assisted coherence distillation is to concentrate the coherence
resource on Bob's side by allowing Alice to perform arbitrary quantum
operations\thinspace \cite{aoc1}. In the asymptotic limit, the collective
operations on many copies of the resource states should be performed to
optimize the rate of coherence distillation. However, the complicated
collective operations are huge challenges even for pure resource states in
experiments. To overcome the difficulty, we consider a new kind of assisted distillation
process by introducing the coordinate local quantum-incoherent operations
and classical communication\thinspace (CoLQICC).

The operation of CoLQICC proposed by us consists of two main parts: i)
Identical local measurements (operations) on Alice's side are \ coordinately
and separably performed on each copy of the resource state. Let the mapping $\Lambda
_{A:\text{CoQ}}$ denote the operation on Alice, for many copies of the
state, there should be $\Lambda _{A:\text{CoQ}}^{\otimes n}\rho
_{_{AB}}^{\otimes n}=\left( \Lambda _{A:\text{CoQ}}\rho _{_{AB}}\right)
^{\otimes n}$. A similar setting can be found in\thinspace \cite{coordin};
ii) Incoherence operations, denoted by the mapping $\Lambda _{B}^{\text{IO}}$,
working on the Bob's side. In our consideration, $\Lambda _{B}^{\text{IO}}$ will collectively
act on the copies of Bob's residual states. Therefore, the
CoLQICC can be described by a complete mapping, i.e., $\Lambda _{\text{CoQ}}\equiv \Lambda
_{B}^{\text{IO}}\circ \Lambda _{A:\text{CoQ}}^{\otimes n}$.

Under the CoLQICC, we define the coordinately assisted distillation of
coherence as:
\begin{equation}
C_{\text{CoQ}}^{A|B}(\rho )=\sup \{\mathcal{R}:\underset{n\rightarrow \infty
}{\lim }\underset{\Lambda _{\text{CoQ}}}{\inf }\Vert \Lambda _{\text{CoQ}%
}(\rho ^{\otimes n})-\Phi _{2}^{\otimes \lfloor n\mathcal{R}\rfloor }\Vert
_{1}=0\},  \label{co}
\end{equation}%
where $\left\Vert O\right\Vert _{1}=\mathrm{Tr}\sqrt{O^{\dag }O}$ is trace
norm. In $D$-dimensional Hilbert space $\mathcal{H}$, the maximal coherent resource
state is $|\Phi _{D}\rangle \equiv\sum_{i=0}^{D-1}|i\rangle /\sqrt{D}$, and $%
\Phi _{2}:=|\Phi _{2}\rangle \langle \Phi _{2}|$ denotes the density matrix
of the 2-dimensional maximal coherent state. The infimum is taken over all
the CoLQICC operators $\Lambda _{\text{CoQ}}$. Obviously, when the state of
Alice and Bob is in a product form, i.e., $\rho _{_{AB}}=\rho _{_{A}}\otimes
\rho _{_{B}}$, the assisted coherence distillation transforms to the
coherence distillation of Bob $C_{d}(\rho _{_{B}})$. In order to facilitate
the experimental study, we further simplify the operations that Alice only
performs\ orthogonal projective measurements, i.e., the measurement
operators satisfy $\mathrm{Tr}(\Xi _{A}^{i}\Xi _{A}^{k})=\delta _{ik}$,$%
~\sum_{i}\Xi _{A}^{i}=\mathbb{I}_{A}$, and $\left( \Xi _{A}^{i}\right)
^{2}=\Xi _{A}^{i}$. In this setting, $\Lambda _{A:\text{CoQ}}^{\otimes
n}\rho _{_{AB}}^{\otimes n}=\sum_{i}(\Xi _{A}^{i}\otimes \mathbb{I}%
_{B})^{\otimes n}\rho _{_{AB}}^{\otimes n}{(\Xi _{A}^{i}\otimes \mathbb{I}%
_{B})}^{\otimes n}$. Thus, in the following sections, we further rewrite the assisted coherence distillation with the definition of the
coordinate local \emph{projective-incoherent} operations and classical communication (CoLPICC).

\textbf{Lemma 1.}---The assisted coherence distillation under the proposed
CoLPICC operations (with the projective measurements coordinately acting on
Alice's side), can be expressed as follows:
\begin{widetext}
\begin{equation}
C_{\text{CoP}}^{A|B}(\rho _{_{AB}})=\underset{\{\Xi _{A}^{i}\}}{\max }%
\,\sum_{i}P_{i}\sup \{R_{i}:\underset{n\rightarrow \infty }{\lim }\underset{%
\{\text{IO}_{B}\}}{\inf }\Vert \Lambda _{B}^{\text{IO}}\left( \rho
_{B}^{i}\right) ^{\otimes n}-\Phi _{2}{}^{\otimes \lfloor nR_{i}\rfloor
}\Vert _{1}=0\},  \label{co-2}
\end{equation}%
\end{widetext}
where $P_{i}$ is the probability distribution and $\rho _{_{B}}^{i}$ is the
residual density with the definitions:
\begin{align}
P_{i}& =\mathrm{Tr}\left( \Xi _{A}^{i}\otimes \mathbb{I}_{_{B}}\rho
_{_{AB}}\right) ,  \notag \\
\rho _{_{B}}^{i}& =\frac{\mathrm{Tr}_{A}\left( \Xi _{A}^{i}\otimes \mathbb{I}%
_{_{B}}\rho _{_{AB}}\right) }{P_{i}}.  \label{Pi-rhoi}
\end{align}%
The maximum is taken over all the Alice's projective measurements and
the infimum is with respect to the optimalization of the incoherent
operations on Bob's side. The rate of the coherence distillation in the
assisted scenario is a probabilistic sum of all the subprocesses, i.e., $%
R=\sum_{i}P_{i}R_{i}$ with the maximum being taken over all the projective
measurements $\{\Xi _{A}^{i}\}$. Finally, the rate of coordinately assisted
coherence distillation becomes $\mathcal{R}=\underset{\Xi _{A}^{i}}{\max }%
\sum_{i}P_{i}R_{i}$ (proof details are shown in Appendix A). Our study
highlights the effect of the local measurements in the auxiliary system,
which was not presented in the conventional definition of the assisted
coherence distillation in\thinspace \thinspace \cite{aoc1}. The measurements
on the auxiliary system cause the coherence distillation process to branch into
several sub-processes, each of which corresponds to a
distillation rate $R_{i}$. On the other hand, since CoLPICC$\subset $CoLQICC$%
\subset $LQICC, one can have the relation $C_{\text{CoP}}^{A|B}(\rho
_{_{AB}})\leq C_{\text{CoQ}}^{A|B}(\rho _{_{AB}})\leq C_{d}^{A|B}(\rho
_{_{AB}})\leq C_{r}^{A|B}(\rho _{_{AB}})$, where the quantum-incoherent
relative entropy $C_{r}^{A|B}(\rho _{_{AB}})=S(\triangle ^{B}\rho
_{_{AB}})-S(\rho _{_{AB}})$ with $\triangle ^{B}(\rho
_{_{AB}}):=\sum_{i}\left( \mathbb{I}_{A}\otimes |i\rangle _{B}\langle
i|\right) \rho \left( \mathbb{I}_{A}\otimes |i\rangle _{B}\langle i|\right) $
and $\mathbb{I}$ being a identity matrix\thinspace \cite{aoc1}. $C_{\text{CoP%
}}^{A|B}$ and $C_{\text{CoQ}}^{A|B}$ correspond to the different sets
CoLPICC and CoLQICC, respectively.

\textbf{Theorem 1.}---With the proposed CoLPICC operators, the coordinately
assisted coherence distillation has an explicit solution:~
\begin{equation}
C_{\text{CoP}}^{A|B}(\rho _{_{AB}})=\underset{\left\{ \Xi _{A}^{i}\right\} }{%
\max }\sum_{i}P_{i}C_{r}\left( \rho _{_{B}}^{i}\right) ,  \label{aver-coh}
\end{equation}%
with the definitions of $P_{i}$ and $\rho _{_{B}}^{i}$ in Eq.\thinspace (\ref%
{Pi-rhoi}) and the projective measurements $\Xi _{A}^{i}$. The measure in
Eq.\thinspace (\ref{aver-coh}) is suitable for the case that $\rho _{_{AB}}$
is a mixed state, and it overcomes the difficulty of huge numbers of copies
in the asymptotic limit and thus is convenient for experimental studies. Our
results also provide an operational interpretation\ of the average relative
entropy of coherence, which should not be simply understood as the one-copy
scenario for pure-state cases used by the authors in\thinspace \cite%
{opticalaoc}, but a more general concept to measure the assisted distillable
coherence even in the asymptotic limit. The proof details of Theorem 1 can
be found in Appendix B, where we first prove the upper bound of the
distillation rate is the average of the relative entropy of coherence. Then
we prove that the upper bound can be achieved by using the typical sequence
technique.

For a pure state density $\Psi _{_{AB}}\equiv \left\vert \Psi
_{_{AB}}\right\rangle \left\langle \Psi _{_{AB}}\right\vert $, through the local
measurements on Alice together with the communications with Bob, any possible
pure decomposition of $\rho _{_{B}}$ can be obtained, i.e., $\rho
_{B}=\sum_{i}p_{i}\Psi _{_{B}}^{i}$ for any set of $\left\{ p_{i}\right\} $
and the corresponding pure state density $\Psi _{_{B}}^{i}\equiv $ $\left\vert \Psi
_{_{B}}^{i}\right\rangle \left\langle \Psi _{_{B}}^{i}\right\vert $.
Therefore,\ based on the definition in Eq.\thinspace (\ref{aver-coh}), we
have $C_{\text{CoP}}^{A|B}(\Psi _{_{AB}})=\underset{\left\{ \Xi
_{A}^{i}\right\} }{\max }\sum_{i}P_{i}C_{r}\left( \Psi _{_{B}}^{i}\right) =%
\underset{\left\{ \Xi _{A}^{i}\right\} }{\max }\sum_{i}P_{i}S\left( \Delta
\Psi _{_{B}}^{i}\right) $, which is identical to the concept of coherence of
assistance (COA) $C_{a}(\rho _{_{B}})$\thinspace \cite{aoc1}. Moreover, one can
find $C_{\text{CoP}}^{A|B}(\Psi _{_{AB}})\leq C_{d}^{A|B}(\Psi
_{_{AB}})=C_{r}^{A|B}(\Psi _{_{AB}})=S(\triangle ^{B}\Psi
_{_{AB}})=S(\triangle \rho _{_{B}})$\thinspace \cite{aoc1}. Now let us
discuss two special cases of pure states: i) The dimension of subsystem $B$ is dim$(%
\mathcal{H}_{B})=2$, then one has $C_{a}(\rho _{_{B}})=$ $S(\triangle \rho
_{_{B}})$\thinspace \cite{aoc1}. Consequently, we have
\begin{equation}
C_{\text{CoP}}^{A|B}( \Psi _{_{AB}})=C_{r}^{A|B}(\Psi _{_{AB}})=S(\triangle \rho _{_{B}}).  \label{Cd-pure}
\end{equation}%
ii) The dimension of auxiliary system (Alice) is dim$(\mathcal{H}_{A})=2$
and that of Bob is dim$(\mathcal{H}_{B})=n$ ($n>2$). For a set of reference
basis $\left\{ \left\vert i\right\rangle \right\} $, on which the quantum
coherence is defined. If the Schmidt decomposition of $\left\vert \Psi
_{AB}\right\rangle $ can be written as follows:

\begin{equation}
|\Psi _{_{AB}}\rangle=\sqrt{\lambda _{1}}|\phi _{A}^{1}\rangle \left(
\sum_{j\neq i}|j\rangle _{B}\right) +\sqrt{\lambda _{2}}|\phi
_{A}^{2}\rangle |i\rangle _{B},  \label{bipartite-special}
\end{equation}%
where $\left\langle \phi _{A}^{2}|\phi _{A}^{1}\right\rangle =0$. Then by
performing the projective measurement of $\{\left( |\phi _{A}^{1}\rangle \pm
|\phi _{A}^{2}\rangle \right) /\sqrt{2}\}$ on Alice, one can easily obtain $C_{\text{%
CoP}}^{A|B}( \Psi _{_{AB}})=S(\triangle \rho
_{_{B}}) $. The expression in Eq.~(\ref{bipartite-special}) also gives an answer
to the remaining issue in\thinspace \cite{aoc1} that for which kind of
high-dimensional pure states, the assisted of coherence (COA), i.e., $C_{%
\text{CoP}}^{A|B}$ for pure states in this work, is equal to the regularized COA for the
infinite copies of the state.

Let us expand to the multipartite-system cases, and take a tripartite pure
state for example. If the Schmidt decomposition of a pure state $|\Psi
\rangle _{_{ABC}}$ with the condition dim$(\mathcal{H}_{A})=2$,\ can be
presented as:
\begin{equation}
|\Psi _{_{ABC}} \rangle=\sqrt{\lambda _{1}}|\phi _{A}^{1}\rangle \left(
\sum_{\left\langle mn|ij\right\rangle =0}|mn\rangle _{BC}\right) +\sqrt{%
\lambda _{2}}|\phi _{A}^{2}\rangle |ij\rangle _{BC},
\label{tripartite-special}
\end{equation}%
where $\left\{ |ij\rangle \right\} $ denotes a set of reference basis and $%
\left\langle \phi _{A}^{2}|\phi _{A}^{1}\right\rangle =0$, we also have the
similar equality in Eq.\thinspace (\ref{Cd-pure}) that:
\begin{equation}
C_{\text{CoP}}^{A|BC}(\left\vert \Psi _{_{ABC}}\right\rangle )=S(\triangle
\rho _{_{BC}}).
\end{equation}
For example, the GHZ-type and $W$-type states satisfy the decomposition in Eq.\thinspace
(\ref{tripartite-special}), thus the above equality holds.

\textit{Assisted coherence distillation in multipartite systems.---} In the
following sections, we will discuss the problems of coordinatedly assisted
coherence distillation in multipartite systems. Let us start from the
tripartite case.

\textbf{Theorem 2.}---In tripartite system, for a pure state $|\Psi
_{_{ABC}}\rangle $ satisfying the condition in Eq.\thinspace (\ref%
{tripartite-special}) and with the dimension of the auxiliary system dim$(%
\mathcal{H}_{A})=2$,\ the following inequality holds,
\begin{equation}
C_{\text{CoP}}^{A|BC}\left( |\Psi _{_{ABC}}\rangle \right) \geq C_{\text{CoP}%
}^{A|B}\left( |\Psi _{_{ABC}}\rangle \right) +C_{\text{CoP}}^{A|C}\left(
|\Psi _{_{ABC}}\rangle \right) ,  \label{ineq1}
\end{equation}

where the first process {\footnotesize $C_{\text{CoP}}^{A|BC}\left( \Psi
_{_{ABC}}\right) =\underset{\left\{ \Xi _{A}^{i}\right\} }{\max }%
\sum_{i}P_{i}C_{r}\left( \rho _{_{BC}}^{i}\right) $} with {\footnotesize $%
\rho _{_{BC}}^{i}=\mathrm{Tr}_{A}\left( \Xi _{A}^{i}\otimes \mathbb{I}%
_{_{BC}}\rho _{_{ABC}}\right) /P_{i}$}, and {\footnotesize $P_{i}=\mathrm{Tr}%
\left( \Xi _{A}^{i}\otimes \mathbb{I}_{_{BC}}\rho _{_{ABC}}\right) $}. The
second process {\footnotesize $C_{\text{CoP}}^{A|B}\left( \Psi
_{_{ABC}}\right) =\underset{\left\{ \Gamma _{A}^{j}\right\} }{\max }%
\sum_{j}P_{j}C_{r}\left( \rho _{_{B}}^{j}\right) $} with {\footnotesize $%
\rho _{_{B}}^{j}= \mathrm{Tr}_{AC}\left( \Gamma _{A}^{j}\otimes \mathbb{I}%
_{_{BC}}\rho _{_{ABC}}\right) /P_{j}$} and {\footnotesize $P_{j}=\mathrm{Tr}%
\left( \Gamma _{A}^{j}\otimes \mathbb{I}_{_{BC}}\rho _{_{ABC}}\right) $}.
The third process {\footnotesize $C_{\text{CoP}}^{A|C}\left( \Psi
_{_{ABC}}\right) =\underset{\left\{ \Theta _{A}^{k}\right\} }{\max }%
\sum_{k}P_{k}C_{r}\left( \rho _{_{C}}^{k}\right) $} with {\footnotesize $%
\rho _{_{C}}^{k}=\mathrm{Tr}_{AB}\left( \Theta _{A}^{j}\otimes \mathbb{I}%
_{_{BC}}\rho _{_{ABC}}\right) /P_{k}$} and {\footnotesize $P_{k}=\mathrm{Tr}%
\left( \Theta _{A}^{k}\otimes \mathbb{I}_{_{BC}}\rho _{_{ABC}}\right) $}.
Obviously, when the state is in a product form, i.e., $|\Psi
_{_{ABC}}\rangle =|\Psi _{_{AB}}\rangle \otimes |\Psi _{_{C}}\rangle $, $%
|\Psi _{_{ABC}}\rangle =|\Psi _{_{AC}}\rangle \otimes |\Psi _{_{B}}\rangle $%
, or $|\Psi_{_{ABC}}\rangle =|\Psi _{_{A}}\rangle \otimes|\Psi _{_{B}}\rangle \otimes |\Psi _{_{C}}\rangle $ the equality holds.

Note that there are actually three optimization processes in the inequality
of Eq.\thinspace (\ref{ineq1}), which are realized by choosing proper projective
measurements $\Xi _{A}^{i}$, $\Gamma _{A}^{j}$, and $\Theta _{A}^{k}$ to
achieve the maximal values of $C_{\text{CoP}}^{A|BC}$, $C_{\text{CoP}}^{A|B}$%
, and $C_{\text{CoP}}^{A|C}$, respectively. The proof of Theorem 2 is shown
in Appendix C. This theorem reveals a distribution formula of the
coordinately assisted distillation of quantum coherence in tripartite
system. The monogamy-like inequality implies that the process of
distillating coherence on the subsystem $BC$ with assistant $A$ cannot be
easily divided into two independent subprocesses, i.e., distillating
coherence on subsystem $B$ or $C$ with assistant $A$. In addition, it also
points out that the assisted coherence distillation occurs between a pair of
$A$ and $B$ (or $C$) must be constrained by the inequality in Eq.\thinspace (%
\ref{ineq1}).

When considering the multipartite case of $N>3$, for a pure state satisfying
the condition by extending Eq.\thinspace (\ref{tripartite-special}) to the
multipartite cases, also with dim$(\mathcal{H}_{A})=2$, then the following
inequality holds:
\begin{align}
C_{\text{CoP}}^{A|B_{1}B_{2}\cdot \cdot \cdot B_{N}}\left( \rho
_{_{AB_{1}B_{2}\cdot \cdot \cdot B_{N}}}\right) \geq \sum_{\alpha =1}^{N}C_{%
\text{CoP}}^{A|B_{\alpha }}\left( \rho _{_{AB_{1}B_{2}\cdot \cdot \cdot
B_{N}}}\right) ,
\end{align}%
where each $C_{\text{CoP}}^{A|B_{\alpha }}$ is obtained by performing the
corresponding optimal measurement $\Xi _{A,\text{opt}}^{\alpha }$ on system $%
A$. The inequality reveals that the assisted distillable coherence of individual subsystems should be constrained by
that of the whole remaining part.

\textbf{Theorem 3}\textit{.}---For a general state $\rho _{AB_{1}B_{2}\cdot
\cdot \cdot B_{N}}$ (either pure or mixed), the following inequality holds:
\begin{equation}
C_{\text{CoP}}^{A|B_{1}\cdot \cdot \cdot B_{N}}(\rho _{_{AB_{1}\cdot \cdot
\cdot B_{N}}})\geq \underset{\{\Xi _{A}^{i}\}}{\max }\sum_{i}P_{i}\left(
\sum_{\alpha =1}^{N}C_{r}(\rho _{B_{\alpha }}^{i})\right) .
\end{equation}%
Note that the inequality above describes that Alice only performs the
optimal measurement $\Xi _{A,\text{opt}}^{i}$ once to achieve the maximal
average\ of the\ sum of the distillable coherence of the residual states
corresponding to each subsystem $B_{\alpha }$. When the state is in a
product form, e.g., $\rho _{_{AB_{1}...B_{N}}}=\rho _{_{AB_{1}}}\otimes \rho
_{_{B_{2}}}...\otimes \rho _{_{B_{N}}}$ (i.e., at most a pair of subsystems
are related) the equality holds. The detailed proof can be found in Appendix
D.
\begin{figure*}[tbp]
\begin{center}
\includegraphics[width=13.5 cm, height=7 cm,clip]{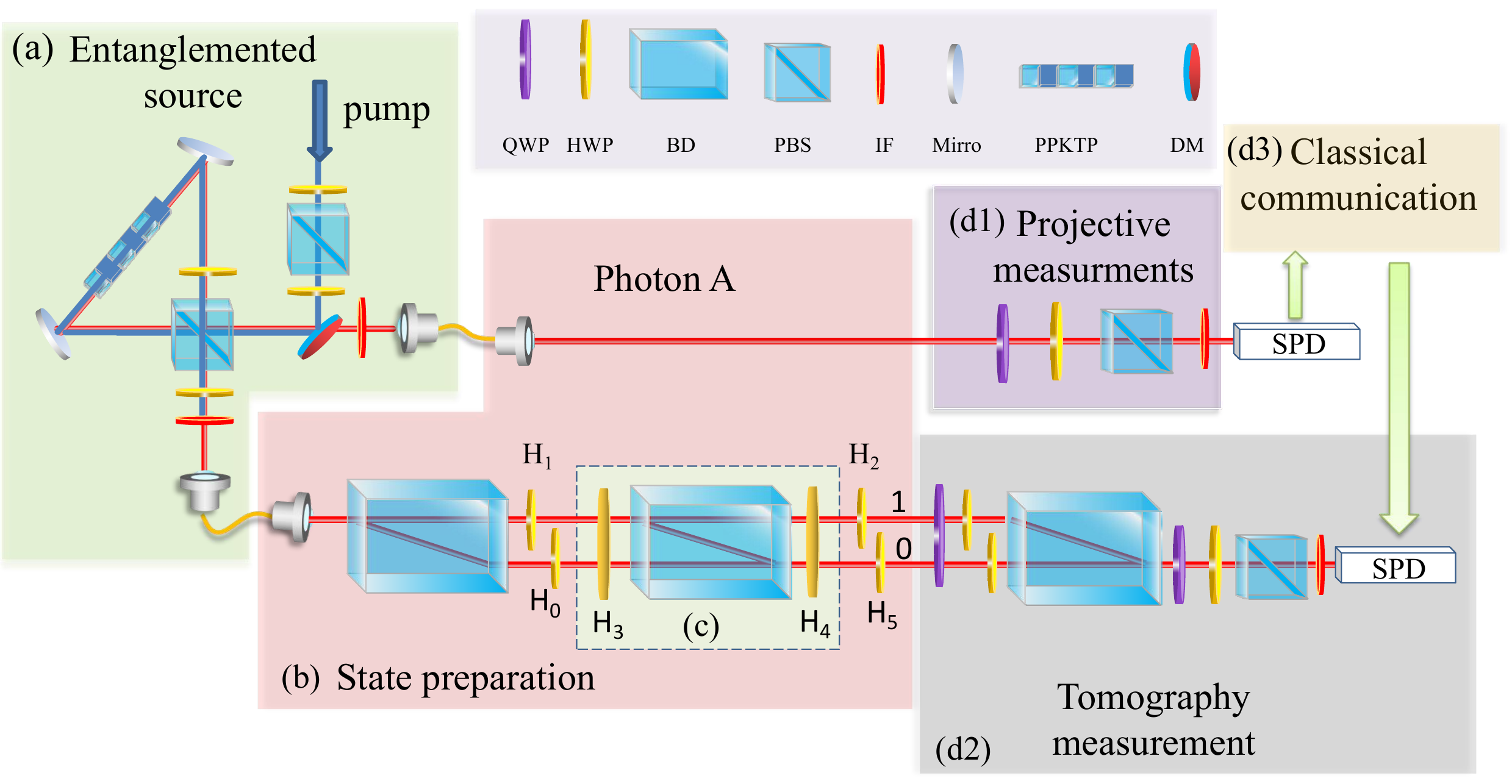} \vspace*{-0.15in}
\end{center}
\caption{~Experimental setups and the stages of the experimental
implementation. \textbf{(a)} The photon pairs with wavelength 810\thinspace
nm were prepared by spontaneous parametric down-conversion\thinspace of the
1.5-cm-long type-II periodically poled potassium titanyl phosphate (PPKTP)
nonlinear crystal. The dichroic mirror (DM) is to reflect photons of 405nm
and transmit the photons of 810nm. \textbf{(b) }Experiment setups for
preparing the state with tripartite quantum correlation. One photon is sent
to the upper experiment installation to act as the auxiliary system $A$. The
other one goes into the lower experimental setting where its polarization
modes interact with the spatial modes. The amplitude decay (AD) and phase
damping (PD) quantum channels are experimentally realized. Different
angles of the half-wave plates (HWP) H$_{1}$ are adjusted to simulate the
superposition coefficient $p$ in the tripartite state in Eq.\thinspace (%
\protect\ref{w-state}) and (\protect\ref{GHZ-state}). The angle of H$_{0}$
is set to zero. \ The angle of H$_{2}$ is set to zero for simulating AD
channel, while set to $\protect\pi /4$ for PD channel. When the angle is set
to $\protect\pi /4$, the HWP can perform the inversions between the
polarization modes $\left\vert H\right\rangle \rightarrow \left\vert
V\right\rangle $ and $\left\vert V\right\rangle \rightarrow \left\vert
H\right\rangle $ (in the text the horizontal ($\left\vert H\right\rangle $)
and vertical ($\left\vert V\right\rangle $) modes are denoted as $\left\vert
0\right\rangle $ and $\left\vert 1\right\rangle $ for simplicity). \textbf{(c)} The
angles of H$_{3}$ and H$_{4}$ are set to $\protect\pi /4$, together with
the beam displacer (BD) between them, to realize an anti-BD, which has
opposite effects of the ordinary BD, i.e., it transmits horizontally
polarized photons and reflects the vertical ones. \textbf{(d1)} Setup for the
projective measurements performed on the subsystem $A$. The
quarter-wave plates (QWP), HWP, and polarizing beam splitters (PBS) are
employed to realize the measurement basises. \textbf{(d2)}
Tomography measurements on the polarization modes of the second photon and
the coupled spatial modes. The residual densities can be constructed based
on the measurement probability of subsystem $A$. The other devices are
interference filters (IF). }
\label{fig:1}
\end{figure*}
Generally, the distribution of assisted coherence distillation in
multipartite systems is difficult to study in experiments. With
the help of the coordinately assisted distillation and the monogamy-like inequalities in Theorem 3 and 4, we can experimentally demonstrate
the distribution relationship based on a linear optical setup.

\section{Experimental demonstration distribution of Coordinately assisted
distillation of coherence \ }

In order to prepare entangled photon pairs, the 405-nm pump laser
(~3 mW) outputs from the continuous laser. The 810-nm photon pairs are
generated by spontaneous parametric down conversion of the 1.5-cm-long
type-II periodically poled potassium titanyl phosphate (PPKTP) nonlinear
crystal in Sagnac loop [shown in the Module (a) in Fig.\thinspace 1].\ The
entangled state is encoded in\ the polarization modes, and thus the
two-qubit space is spanned by the basis vectors $\left\{ |i\rangle
_{A}|j\rangle _{B}\right\} $ with $i$, $j=0$, $1$. We obtain 45000/s
entangled photon pairs with the concurrence being $0.982$, and the fidelity to the maximally entangled pure state $\left( \left\vert
11\right\rangle _{AB}+\left\vert 00\right\rangle _{AB}\right) /\sqrt{2}$,
reaching $99.8\%$. In the Module (b) of Fig.\thinspace 1, by using the beam
displacer (BD), the polarization modes of photon B ($|j\rangle _{B}$) is
coupled to the spatial modes ($|k\rangle _{C}$ with $k=0$, $1$). Based on
the polarization-spatial interactions, we prepare the tripartite
states\thinspace \cite{exptri}. Moreover, in this work, two types of quantum
channels are constructed to realize the polarization-spatial
interactions, one is the depolarization (PD) channel corresponding to the
following map:
\begin{align}
& |0\rangle _{B}|0\rangle _{C}\rightarrow |0\rangle _{B}|0\rangle _{C}{},
\notag \\
& |1\rangle _{B}|0\rangle _{C}\rightarrow \sqrt{1-p}|1\rangle _{B}|0\rangle
_{C}+\sqrt{1-p}|0\rangle _{B}|1\rangle _{C},  \label{PD}
\end{align}%
and the other is the amplitude (AD)\thinspace\ channel:
\begin{align}
& |0\rangle _{B}|0\rangle _{C}\rightarrow |0\rangle _{B}|0\rangle _{C}{},
\notag \\
& |1\rangle _{B}|0\rangle _{C}\rightarrow \sqrt{1-p}|1\rangle _{B}|0\rangle
_{C}+\sqrt{1-p}|1\rangle _{B}|1\rangle _{C}.  \label{AD}
\end{align}%
With the help of the two channels above, we prepare two types of three-qubit
entangled states\thinspace \cite{exptri}. For the initial state $\frac{1}{%
\sqrt{3}}\left( \left\vert 10\right\rangle _{AB}+\sqrt{2}\left\vert
01\right\rangle _{AB}\right) \left\vert 0\right\rangle _{C}$, the AD channel
produces the $W$-type state
\begin{equation}
|\phi \rangle =\frac{1}{\sqrt{3}}|100\rangle +\sqrt{\frac{2}{3}}\left( \sqrt{%
1-p}|010\rangle +\sqrt{p}|001\rangle \right) .  \label{w-state}
\end{equation}%
For $p=1/2$, the state becomes the $W$ state. The subscripts $A$, $B$, $C$
are omitted for simplicity.\ In the experiment, the parameter $p$ can be
simulated by the rotation angle $\theta $ of HWP$_{1}$ with the relation $%
p=\sin ^{2}(2\theta )$.

For the initial state $\frac{1}{\sqrt{2}}\left( \left\vert 11\right\rangle
_{AB}+\left\vert 00\right\rangle _{AB}\right) \left\vert 0\right\rangle _{C}$%
, the PD channel produces the GHZ-type state
\begin{equation}
|\phi \rangle =\frac{1}{\sqrt{2}}(|000\rangle +\sqrt{1-p}|110\rangle +\sqrt{p%
}|111\rangle ),  \label{GHZ-state}
\end{equation}%
which becomes the GHZ state for $p=1$. In the following section, we
experimentally study the assisted coherence distillation and verify the
inequalities (\ref{ineq1}) based on the prepared tripartite entangled states.
\begin{figure}[tbp]
\epsfig{file=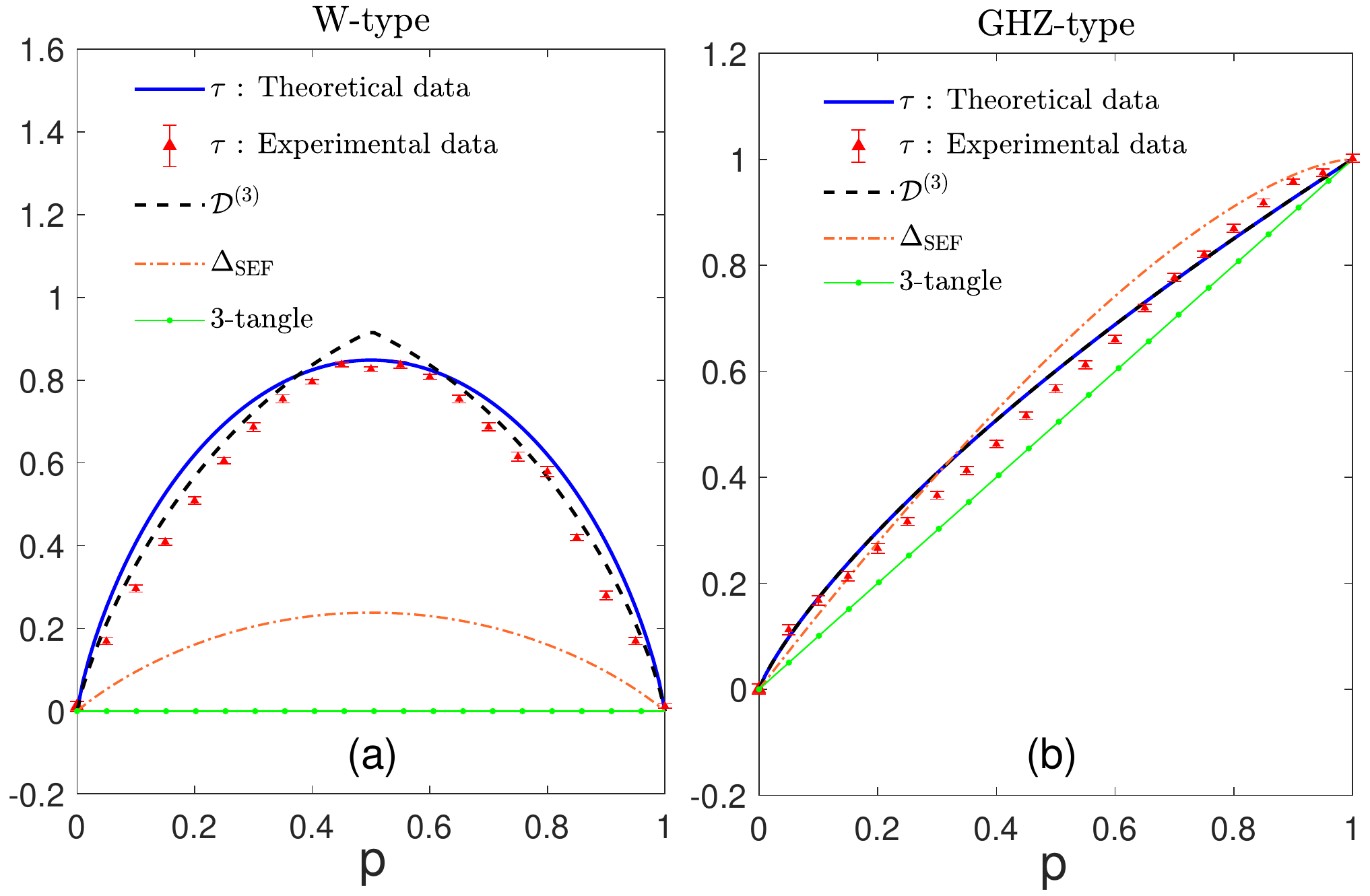,width=8.9cm, height=6cm}
\caption{(colour online)Experimental and theoretical results of the distribution of the
coordinately assisted distillation of quantum coherence in the tripartite
system. Three measures of multipartite correlations are considered. The blue
solid line is the theoretical curve of the distribution core $\protect\tau %
\equiv $ $C_{\text{CoP}}^{A|BC}-C_{\text{CoP}}^{A|B}-C_{\text{CoP}}^{A|C}$
of the assisted coherence distillation\ defined from the inequality in
Eq.\thinspace (\protect\ref{ineq1}). The red triangle-errorbar denotes the
experimental result of $\protect\tau $. The black dashed line displays the
genuine quantum correlation based on the multipartite discord $\mathcal{D}%
^{(3)}$\thinspace \protect\cite{genuine-discord}. The orange dot-dashed line
describes multipartite entanglement indicator based on the difference of the
squared entanglement of formation $\Delta _{\text{SEF}}\,$\protect\cite%
{multi-formation-Bai}. The green dot-solid line denotes the
three-tangle\thinspace \protect\cite{exptri}.}
\label{fig:2}
\end{figure}

\emph{Experimental results}.---In the experiment, we perform optimal
projective measurement on subsystem $A$ to obtain the assisted coherence
distillation $C_{\text{CoP}}^{A|BC}$ in Eq.\thinspace (\ref{aver-coh}). One
can find that the optimal measurement basises should be $\left( |0\rangle
\pm |1\rangle \right) /\sqrt{2}$, which is due to that both GHZ-type and $W$%
-type states satisfy the Schmidt decomposition in Eq.\thinspace (\ref%
{tripartite-special}). Then by doing tomograph, the residual density matrix $%
\rho _{_{BC}}$ (corresponding to the measurement probability on $A$)\ can be
obtained.

To $C_{\text{CoP}}^{A|B}$ and $C_{\text{CoP}}^{A|C}$, one should take into
account the reduced density $\rho _{_{AB}}$and $\rho _{_{AC}}$. In order
to find the optimal measurement on $A$,\ we introduce a general set of
projective measurement basises denoted by $\cos \theta |0\rangle \pm \sin
\theta e^{i\varphi }|1\rangle $. First, let us study the initial $W$-type
state, another measure, i.e., $l_{1}$ norm of coherence\thinspace \cite%
{quantify}, is employed to facilitate the analysis of the maximal coherence
in the residual density. By numerically calculation, we find the behavior of
the$\ l_{1}$ norm of coherence is similar with the relative entropy of
coherence in the considered state. More importantly, $l_{1}$ norm of
coherence has a simple definition, and thus one can easily obtain the
average $l_{1}$ norm of coherence of the subsystems $B$ and $C$ and which is
found to be proportional to $\sqrt{1-p}\sin \theta \cos \theta $. Obviously,
the measurement of $\theta =\pi /4$ (i.e., the measurement basises $\left(
|0\rangle \pm |1\rangle \right) /\sqrt{2}$) is optimal to help the system $B$%
\thinspace ($C$) to\ capture the maximal average coherence. While, for the
initial GHZ-type state, we found the behaviors of the $l_{1}$ norm of
coherence and relative entropy of coherence are different. Fortunately, the
simple structure of GHZ-type state makes it possible to analyse the relative entropy.
The detailed calculations are shown in the Appendix.
We find that the optimal measurement basises of $A$ are $\left( |0\rangle \pm
|1\rangle \right) /\sqrt{2}$ to obtain $C_{\text{CoP}}^{A|B}$, while $\left(
|0\rangle \text{, }|1\rangle \right) $ to obtain $C_{\text{CoP}}^{A|C} $.

In Fig.\thinspace 2(a), we prepare the $W$-type tripartite state, and
perform the optimal measurement on photon $A$. Then the residual states of
the subsystem $BC$, $B$, and $C$ can be detected by tomography. Furthermore,
one obtains the coherence of the residual states, and thus the assisted
distillable coherence, i.e., $C_{\text{CoP}}^{A|BC}$, $C_{\text{CoP}}^{A|B}$%
, and$\ C_{\text{CoP}}^{A|C}$. We define the distribution
core $\tau \equiv $ $C_{\text{CoP}}^{A|BC}-C_{\text{CoP}}^{A|B}-C_{\text{CoP}%
}^{A|C}$ \ and show its theoretical and experimental results versus the
superposition parameter $p$ in Fig.\thinspace 2(a). One can find that $\tau
\geq 0$ in the whole parameter region, which verifies the inequality\thinspace
(\ref{ineq1}). Moreover, $\tau $ reaches its maximum at $p=1/2$, where the
tripartite state becomes the $W$ state, i.e., $|\phi \rangle _{_{W}}=\left(
|100\rangle +|010\rangle +|001\rangle \right) /\sqrt{3}$. While, $\tau $
reaches zero at $p=0$ and $1$, where the tripartite quantum correlation
degenerates into the bipartite correlation. Certainly, if one chooses
subsystem $B$ or $C$ as the auxiliary system, the process of the assisted
coherence distillation is different. The values of the distribution core $\tau _{B\text{ }}$or $%
\tau _{C}$ may not be zero (with the subscripts $B$, $C$ corresponding to
the auxiliary system). Then the symmetrized form $\tau _{m}=\min (\tau $, $%
\tau _{B}$, $\tau _{C})$ should be introduced. In this paper, we only
experimentally investigate $\tau$ (with subsystem $A$ being the auxiliary)

It is known that $W$-type state is rich in genuine tripartite quantum
correlations\thinspace \cite{multi-formation-Bai,genuine-discord}. We
believe that the nonzero values of the core $\tau $ which characterizes the
distribution rule of the assisted distillable coherence, are in close
relationship with the genuine quantum correlations. We numerically calculate
the genuine tripartite quantum entanglement $\Delta _{\text{SEF}}$\thinspace
\cite{multi-formation-Bai} and the genuine tripartite quantum discord $%
\mathcal{D}^{(3)}$ \cite{genuine-discord}, whose definition can be found in
the appendix. One can find the similar behaviors in $\tau $, $\Delta _{\text{%
SEF}}$, and $\mathcal{D}^{(3)}$, e.g., their zero values appear at $p=0$, $1$%
, and the maximal values reach at $p=1/2$. The increase (decrease) of $\tau $
is synchronized with\ the increase (decrease) of $\Delta _{\text{SEF}}$ and $%
\mathcal{D}^{(3)}$. We also consider another well-known measure, i.e., the
three-tangle\thinspace \thinspace \cite{exptri}, which is found to be always
zero in the considered region of $p$. It implies that nonzero $\tau $ should
be connected with the multipartite correlation that cannot be detected by
three-tangle but can be characterized by $\Delta _{\text{SEF}}$ and $%
\mathcal{D}^{(3)}$.

In Fig.\thinspace 2(b), the case of GHZ-type states is studied. One can see
that $\tau $, $\Delta _{\text{SEF}}$, $\mathcal{D}^{(3)}$, and three-tangle
all increase monotonously as $p$ increases, which is quite different from
that in the case of $W$-type states. More specially, $\tau $ and $\mathcal{D}%
^{(3)}$ are completely coincident. The zero values of the four quantities
are found at $p=0$, where the genuine tripartite correlation disappears,
instead, only bipartite correlation exists. While, at $p=1$, the state
becomes GHZ state, which displays the maximal genuine tripartite
correlation, and then $\tau $ also reaches its maximum. \

\section{Conclusion}

We have considered the issue of assisted coherence distillation in the
asymptotic limit. Different types of measurements on the auxiliary
system were discussed. Then, we focused on coordinatedly performing projective measurements on the auxiliary of each resource state copy.
Our study highlights the effects of the auxiliary's measurements, which was not taken seriously in the conventional scenario of the
assisted coherence distillation. The measurements on the auxiliary system
causes the coherence distillation process to be branched into several subprocesses, each of
which corresponds to its own distillation rate. Finally, a simple formula of the
assisted distillable coherence is obtained as the maximal average coherence
of the residual states, which is also applicable for the cases that the
considered system and its auxiliary are in a composite mixed state. The
formula provides a possible way for the experimental research of the assisted
coherence distillation.

We for the first time investigated the assisted coherence distillation in
multipartite systems. Monogamy-like inequalities were given to constrain the
distribution of the assisted distillable coherence in the subsystems. We
experimentally prepared two types of tripartite correlated states, i.e., the
$W$-type and GHZ-type states, and experimentally study the assisted
coherence distillation. Theoretical and experimental results agree well to
verify our distribution inequalities. Three types of measures of
multipartite correlation were also considered. Our results reveal that the assisted coherence distillation is in close relationship with the genuine
multipartite quantum correlations which sometimes cannot even be detected by the
well-known measure---three-tangle, e.g., in the $W$-type states.

\begin{center}
\textbf{ACKNOWLEDGMENTS}
\end{center}

The work is supported by the National Natural Science Foundation of China
(NSFC) (11775065, 12175052, 11875120, 62105086); the NKRDP of China (Grant No.
2016YFA0301802).

\subsection*{\textbf{APPENDIX A: PROOF OF LEMMA 1}}

In this appendix, we will show the proof of Lemma 1. Based on our proposed
CoLQICC, after coordinately performing the projective measurements on the $n$
copies of the resource state, we obtain a mixed form of Alice's
post-measurement state and Bob's residual state:
\begin{equation}
\Lambda _{A:\text{CoP}}\left( \rho _{_{AB}}^{\otimes n}\right)
=\sum_{i}P_{i}(\Xi _{A}^{i}\otimes \rho _{B}^{i})^{\otimes n},
\end{equation}%
with $P_{i}=\mathrm{Tr}\left( \Xi _{A}^{i}\otimes \mathbb{I}_{_{B}}\rho
_{_{AB}}\right) $, and $\rho _{_{B}}^{i}=\mathrm{Tr}_{A}\left( \Xi
_{A}^{i}\otimes \mathbb{I}_{_{B}}\rho _{_{AB}}\right) /P_{i}$. Recalling the
map $\Lambda _{\text{CoP}}\equiv \Lambda _{B}^{\text{IO}}\circ \Lambda _{A:%
\text{CoP}}$ and substituting the equation above into the definition of the
coordinately assisted coherence distillation, we have:
\begin{widetext}
\begin{eqnarray}
C_{\text{CoP}}^{A|B}(\rho ) &=&\sup \{\mathcal{R}:\underset{n\rightarrow
\infty \,}{\lim }\underset{\Lambda _{\text{CoP}}}{\inf }\Vert \Lambda _{%
\text{CoP}}(\rho _{_{AB}}^{\otimes n})-\Phi _{2}^{\otimes \lfloor n\mathcal{R%
}\rfloor }\Vert _{1}=0\}  \notag \\
&=&\max_{\Xi _{A}^{i}}\sup \{\mathcal{R}:\underset{n\rightarrow \infty }{%
\lim \,}\underset{\text{IO}_{B}}{\inf }\Vert \sum_{i}P_{i}\left( \Xi
_{A}^{i}\right) ^{\otimes n}\otimes \Lambda _{B}^{\text{IO}}\left( \rho
_{B}^{i}\right) ^{\otimes n}-\Phi _{2}^{\otimes \lfloor n\mathcal{R}\rfloor
}\Vert _{1}=0\}.
\end{eqnarray}%
\end{widetext}
We find that after the projective measurements, incoherent operations are
only performed on Bob's side to finally realize the goal of the coherence
distillation. Therefore, focusing on the core part, the trace norm actually
becomes:
\begin{widetext}
\begin{align}
& ~~~~D^{\Xi _{A}^{i}}\left( \rho _{B}^{i}\right) \equiv\Vert \sum_{i}P_{i}(\Xi
_{A}^{i})^{\otimes n}\otimes \Lambda _{B}^{\text{IO}}(\rho
_{B}^{i})^{\otimes n}-\sum_{i}P_{i}(\Xi _{A}^{i})^{\otimes n}\otimes \Phi
_{2}{}^{\otimes \left\lfloor nR_{i}\right\rfloor }\Vert {}_{1}  \notag \\
& \leq \sum_{i}P_{i}\Vert \left( \Xi _{A}^{i}\right) ^{\otimes n}\otimes
\Lambda _{B}^{\text{IO}}(\rho _{B}^{i})^{\otimes n}-\left( \Xi
_{A}^{i}\right) ^{\otimes n}\otimes \Phi _{2}{}^{\otimes \left\lfloor
nR_{i}\right\rfloor }\Vert {}_{1}  \notag \\
& =\sum_{i}P_{i}\Vert \Lambda _{B}^{\text{IO}}(\rho _{B}^{i})^{\otimes
n}-(\Phi _{2})^{\otimes \left\lfloor nR_{i}\right\rfloor }\Vert _{1{}},
\end{align}%
\end{widetext}
where the first inequality is due to the convexity of trace norm, and the
second equality comes from the fact $\left\Vert \Xi \otimes M\right\Vert
_{1}=\left\Vert M\right\Vert _{1}$ for a hermitian matrix $M$ and a matrix $%
\Xi $ of rank 1. Now recalling the original concept of coherence
distillation in\thinspace \cite{wintedistill}, when $n\rightarrow \infty $,
proper incoherent operations on Bob can be found to make $(\rho
_{B}^{i})^{\otimes n}$ approach $(\Phi _{2})^{\otimes \left\lfloor
nR_{i}\right\rfloor }$\ asymptotically, i.e., existing an arbitrarily small $%
\varepsilon _{i}{}\rightarrow 0$ that the trace norm satisfies
\begin{equation}
\underset{\text{IO}_{B}}{\inf }\Vert \Lambda _{B}^{\text{IO}}(\rho
_{B}^{i})^{\otimes n}-(\Phi _{2})^{\otimes \left\lfloor nR_{i}\right\rfloor
}\Vert _{1{}}\leq \varepsilon _{i}{}.
\end{equation}%
Then one has
\begin{equation}
\underset{n\rightarrow \infty }{\lim }\underset{\,\,\text{IO}_{B}}{\,\inf }%
D^{\Xi _{A}^{i}}\left( \rho _{B}^{i}\right) \leq \,\varepsilon \equiv
\underset{n\rightarrow \infty }{\lim }\sum_{i}P_{i}\varepsilon
_{i}{}\rightarrow 0,
\end{equation}
which implies that the process of the coordinately assisted coherence
distillation, i.e., the asymptotic incoherent transformation $\rho
_{_{AB}}^{\otimes n}\overset{\text{CoLPICC}}{\mapsto }\overset{1-\varepsilon
}{\approx }\Phi _{2}^{\otimes \lfloor n\mathcal{R}\rfloor }$ is achievable
as $n\rightarrow \infty $, $\varepsilon \rightarrow 0$. Subsequently, the
rate of coherence distillation in this assisted scenario is a probabilistic
sum of all the parts: $R=\sum_{i}P_{i}R_{i}$, whose maximum is taken over
all the projective measurements $\{\Xi _{A}^{i}\}$. Finally, the rate of
coordinately assisted distillation of coherence becomes $\mathcal{R}=%
\underset{\Xi _{A}^{i}}{\max }\sum_{i}P_{i}R_{i}$.

\subsection*{\textbf{APPENDIX B: PROOF OF THEOREM 1}}

First let us prove the upper bound of the rate of the coordinately assisted
coherence distillation, i.e., $\mathcal{R\leq }\underset{\{\Xi _{A}^{i}\}}{%
\max }\sum_{i}P_{i}C_{r}(\rho _{B}^{i})$.

Due to the continuity of the entropy, for two states $\rho _{_{AB}},~\sigma
_{_{AB}}$, supposing the trace norm satisfies $\Vert \rho _{_{AB}}-\sigma
_{_{AB}}\Vert _{1}\leq \varepsilon $ (with $0\leq \varepsilon \leq 1/2$), the
QI relative entropy (i.e., the relative entropy between a state and a
quantum-incoherent state) is proved to be continuous \cite{aoc1}, i.e.,
\begin{equation*}
|C_{r}^{A|B}(\rho _{_{AB}})-C_{r}^{A|B}(\sigma _{_{AB}})|\leq \varepsilon
\log _{2}d_{AB}+2h(\varepsilon /2),
\end{equation*}%
where the QI relative entropy $C_{r}^{A|B}(\rho _{_{AB}})\equiv S(\Delta
^{B}\rho _{_{AB}})-S(\rho _{_{AB}})$ with $S(\rho )$ being the von Neumann
entropy, and\ the function~$h(x)\equiv -x\log _{2}(x)-(1-x)\log _{2}(1-x)$,
and $d_{AB}$ is the dimension of the Hilbert space. For a projective
measurement $\left\{ \Xi _{A}^{i}\right\} $ on Alice, when we have the trace
norm $\Vert \sum_{i}P_{i}\Lambda _{B}^{\text{IO}}(\Xi _{A}^{i}\otimes \rho
_{B}^{i})^{\otimes n}-\sum_{i}P_{i}\left( \Xi _{A}^{i}\right) ^{\otimes
n}\otimes \Phi _{2}{}^{\otimes nR_{i}}\Vert _{1}\leq \varepsilon $ at the
limit of $n\rightarrow \infty $ and taking the infimum over $\Lambda _{B}^{%
\text{IO}}$, one can obtain\bigskip\ the asymptotic continuity
\begin{widetext}
\begin{equation}
~~~~C_{r}^{A|B}\left[ \sum_{i}P_{i}\Lambda _{B}^{\text{IO}}(\Xi
_{A}^{i}\otimes \rho _{B}^{i})^{\otimes n}\right] \geq C_{r}^{A|B}\left[
\sum_{i}P_{i}\left( \Xi _{A}^{i}\right) ^{\otimes n}\otimes \Phi
_{2}{}^{\otimes nR_{i}}\right] -f(\varepsilon ),
\end{equation}%
\end{widetext}
where $f(\varepsilon )\equiv n\varepsilon \log
_{2}d_{AB}+2h(\varepsilon /2){}$. The right hand side (RHS) of the
inequality
\begin{widetext}
\begin{align}
\text{RHS}& =S\left[ \sum_{i}P_{i}\left( \Xi _{A}^{i}\right) ^{\otimes
n}\otimes \Delta \Phi _{2}{}^{\otimes nR_{i}}\right] -S\left[
\sum_{i}P_{i}\left( \Xi _{A}^{i}\right) ^{\otimes n}\otimes \Phi
_{2}{}^{\otimes nR_{i}}\right] -f(\varepsilon )  \notag \\
& =\sum_{i}P_{i}S(\triangle \Phi _{2}^{\otimes
nR_{i}})+H\{P_{i}\}-\sum_{i}P_{i}S(\Phi _{2}^{\otimes
nR_{i}})-H\{P_{i}\}{}-f(\varepsilon ){}  \notag \\
& =\sum_{i}nP_{i}R_{i}S(\triangle \Phi _{2})-f(\varepsilon )  \notag \\
& =n\sum_{i}P_{i}R_{i}-f(\varepsilon ),
\end{align}
\end{widetext}
In the second equality, we make use of the property of von
Neumann entropy, i.e., $S\left[ \sum_{i}p_{i}\left\vert i\right\rangle
\left\langle i\right\vert \otimes \rho _{i}\right] =H(p_{i})+\sum_{i}p_{i}S(%
\rho _{i})$ where $H(p_{i})$ is the Shannon entropy and $\left\vert
i\right\rangle $ are orthogonal states.\ When $n\rightarrow \infty $, there
is $\varepsilon \rightarrow 0$. In addition, since the relative entropy
cannot be increased by the incoherent operations, one has
\begin{widetext}
\begin{equation}
~~~~C_{r}^{A|B}[\sum_{i}P_{i}(\Xi _{A}^{i}\otimes \rho _{B}^{i})^{\otimes
n}]{}\geq C_{r}^{A|B}\left[ \sum_{i}P_{i}\Lambda _{B}^{\text{IO}}(\Xi
_{A}^{i}\otimes \rho _{B}^{i})^{\otimes n}\right] \geq n\sum_{i}P_{i}R_{i}.{}
\end{equation}%
\end{widetext}Based on the definition of the relative entropy in terms of
entropy, one can easily have $C_{r}^{A|B}[\sum_{i}P_{i}(\Xi _{A}^{i}\otimes
\rho _{B}^{i})^{\otimes n}]=n\sum_{i}P_{i}C_{r}\left( \rho _{B}^{i}\right) $%
. Thus, the upper bound is given in the inequality
\begin{equation}
R=\sum_{i}P_{i}R_{i}\leq \sum_{i}P_{i}C_{r}(\rho _{B}^{i}).
\end{equation}%
Then one can obtain the maximum of $R$ by taking all the projective
measurement $\{\Xi _{A}^{i}\}$, i.e., $\mathcal{R}=\underset{\{\Xi _{A}^{i}\}%
}{\max }\sum_{i}P_{i}R_{i}\leq \underset{\{\Xi _{A}^{i}\}}{\max }%
\sum_{i}P_{i}C_{r}(\rho _{B}^{i})$.

Now, we should prove that the upper bound of the distillation rate can be
achieved. The typicality technique will be employed to analysize the
asymptotic limit case\thinspace \cite{wintedistill,wintertypical}. Let us
start from the purification of $\rho _{AB}$, i.e., $\rho _{AB}\underset{%
\text{purification}}{\Longrightarrow }|\Psi _{ABE}\rangle \langle \Psi
_{ABE}|$. We suppose that the optimal projective measurement performed on
system $A$ is $\left\{ \Pi _{A}^{\nu }\right\} $, and Alice sends the
outcomes to Bob by a classical way. Then the post-measurement state,
corresponding to the projector $\Pi _{A}^{\nu }\equiv \left\vert \Pi
_{A}^{\nu }\right\rangle \left\langle \Pi _{A}^{\nu }\right\vert $, becomes
proportional to\ $|\Pi _{A}^{\nu }\rangle |\psi _{BE}^{\nu }\rangle $ with
the probability $P_{\nu }=\mathrm{Tr}\left( \Pi _{A}^{\nu }\otimes \mathbb{I}%
_{BE}\rho _{_{ABE}}\right) $. Then, \bigskip to the $n$ copies of $|\Psi
_{ABE}\rangle $, after coordinately and independently performing the
projective measurement $\left( \Pi _{A}^{\nu }\right) ^{\otimes n}$, the
post-measurement state will be proportional to
\begin{equation}
|\Phi _{ABE}^{\nu }\rangle _{{}}^{\otimes n}\sim |\Pi _{A}^{\nu }\rangle
^{\otimes n}|\psi _{BE}^{\nu }\rangle ^{\otimes n}.  \label{pt-state}
\end{equation}%
Then, if we implement the type measurement $M_{P}$ on the subsystem $B$,
i.e.,
\begin{equation}
M_{P}=\sum_{i^{n}\in T_{n}^{B,\nu }(P)}\left\vert i_{\nu }^{n}\right\rangle
\left\langle i_{\nu }^{n}\right\vert ,
\end{equation}%
where $\left\vert i_{\nu}^{n}\right\rangle \equiv \left\vert
i_{1},i_{2},...,i_{n}\right\rangle _{\nu}$ describes the typical state
sequence corresponding to the space of the post-measurement states.\ Each
group $\left\{ \left\vert i_{n}\right\rangle \right\} $, corresponding to
the $n$th copy, can be the reference basis on which the quantum coherence is
defined. The type measurement $M_{P}$, consisting of the projectors, can
help us to choose all the typical sequences corresponding to the probability
distribution $P$, which derives from the considered state. Thus, $P$ is
used\ to represent the type of strings $\left\vert i^{n}\right\rangle $ with
length $n$. $T_{n}^{B,\nu }\left( P\right) $ denotes the type class of $P$
corresponding to the measurement $\Pi _{A}^{\nu }$, then $\delta $-typical ($%
\delta >0$) class satisfies \
\begin{equation}
T_{n}^{B,\nu }\left( P\right) =\left\{ \left\vert i_{\nu }^{n}\right\rangle
:\left\vert -\frac{1}{n}\log p_{i_{\nu }^{n}}-H\left( P\right) \right\vert
<\delta \right\} ,  \label{typedefined}
\end{equation}%
where the probability sequence $p_{i_{\nu}^{n}}=p_{i_{1}}^{\nu}p_{i_{2}}^{%
\nu}\cdots p_{i_{n}}^{\nu}$, with the definition $p_{i}=\left\langle
i\right\vert \rho _{B}^{\nu }\left\vert i\right\rangle $ for each set of
basis $\left\{ i_{n}\right\} $ and $\rho _{B}^{\nu }$ being the reduced
density matrix of system $B$ after the measurement $\Pi _{A}^{\nu }$. The
Shannon entropy $H\left( P\right) =-\sum\nolimits_{i}p_{i}\log p_{i}$. The
length of the $\delta $-typical class $\left\vert T_{n}^{B,\nu
}(P)\right\vert $ should be
\begin{equation}
2^{n(S\left( \Delta \rho _{B}^{\nu }\right) -\delta )}\leq \left\vert
T_{n}^{B,\nu }(P)\right\vert \leq 2^{n(S\left( \Delta \rho _{B}^{\nu
}\right) +\delta )},
\end{equation}%
i.e., $\left\vert T_{n}^{B,\nu }(P)\right\vert $ indicates the number of the
typical sequences, and $\Delta (\rho _{B}^{\nu })=\sum_{i}\left\vert
i\right\rangle \left\langle i\right\vert \rho _{B}^{\nu }\left\vert
i\right\rangle \left\langle i\right\vert $ with $\left\vert i\right\rangle $
being the reference basis vector in each copy, on which the quantum
coherence is defined. Then the dimension of the typical space holds
\begin{equation}
2^{n(S\left( \Delta \rho _{B}^{\nu }\right) -\delta )}\leq \dim \left[
T_{n}^{B,\nu }(P)\right] \leq 2^{n(S\left( \Delta \rho _{B}^{\nu }\right)
+\delta )}.
\end{equation}%
After the measurement $\Pi _{A}^{\nu }$ and the type measurement, the state
of $B$ and $E$ can be expressed as
\begin{equation}
|\Phi _{BE}^{\nu }\rangle _{T_{n}}^{\otimes n}=\frac{1}{\sqrt{\left\vert
T_{n}^{B,\nu }(P)\right\vert }}\sum_{i^{n}\in T_{n}^{B,\nu }(P)}\left\vert
i_{\nu }^{n}\right\rangle \left\vert \varphi _{E,\nu }^{i^{n}}\right\rangle .
\end{equation}%
Due to the typical subspace theorem\thinspace \cite{wintertypical}, we can
divided the type $P$ into two types $F$ and $M$ corresponding to the subsets
$\left\{ f\right\} $ and $\left\{ m\right\} $ that $\left\vert T_{n}^{B,\nu
}(P)\right\vert =\left\vert F_{\nu }\right\vert \cdot \left\vert M_{\nu
}\right\vert $. Since the property of the entropy
\begin{align}
&S\left( \Delta \rho _{B}^{\nu }\right) {}\notag\\
&=S\left[ \Delta ^{B}\left( |\psi
_{BE}^{\nu}\rangle \left\langle \psi _{BE}^{\nu}\right\vert \right) \right]{}\notag\\
&=I_{B:E}\left[ \Delta ^{B}\left( |\psi _{BE}^{\nu}\rangle \left\langle \psi
_{BE}^{\nu}\right\vert \right) \right] +S_{B|E}\left[ \Delta ^{B}\left( |\psi
_{BE}^{\nu}\rangle \left\langle \psi _{BE}^{\nu}\right\vert \right) \right] ,{}\notag
\end{align}%
where the post-measurement state $|\psi _{BE}^{\nu }\rangle $
comes from Eq.\thinspace (\ref{pt-state}), and $I_{B:E}$ denotes the mutual
information and $S_{B|E}$ is the conditional entropy. Then the length of the
subset $\left\{ m\right\} $ is $\left\vert M_{\nu }\right\vert \approx
2^{nI_{B:E}}.$

By using the Schmidt decomposition form of $|\psi _{BE}^{\nu}\rangle $, one
can simply obtain
\begin{equation}
\Delta ^{B}\left( |\psi _{BE}^{\nu}\rangle \left\langle \psi
_{BE}^{\nu}\right\vert \right) =\sum_{i}q_{i}^{\nu}\left\vert i\right\rangle
_{B}\left\langle i\right\vert \otimes \left\vert \varphi
_{i}^{\nu}\right\rangle _{E}\left\langle \varphi _{i}^{\nu}\right\vert ,{}
\end{equation}%
where $q_{i}^{\nu}=\sum_{k}\left( \lambda _{k}^{\nu}\right) ^{2}\left\vert
\left\langle i|\phi _{k}\right\rangle \right\vert ^{2}$ with $\lambda
_{k}^{\nu}$ being the Schmidt coefficient and $\left\vert \phi
_{k}\right\rangle $ is the Schmidt basis of $B$. Then the mutual information%
\begin{small}
\begin{align}
I_{B:E}\left[ \Delta ^{B}\left( |\psi _{BE}^{\nu}\rangle \left\langle \psi
_{BE}^{\nu}\right\vert \right) \right] & =S\left( \sum_{i}q_{i}^{\nu}\left\vert
i\right\rangle \left\langle i\right\vert \right) +S\left(
\sum_{i}q_{i}^{\nu}\left\vert \varphi _{i}^{\nu}\right\rangle \left\langle
\varphi _{i}^{\nu}\right\vert \right){}\notag\\
&-S\left( \sum_{i}q_{i}^{\nu}\left\vert i\right\rangle \left\langle
i\right\vert \otimes \left\vert \varphi _{i}^{\nu}\right\rangle \left\langle
\varphi _{i}^{\nu}\right\vert \right) {}  \notag \\
& =S\left( \rho _{E}^{\nu}\right) {}=S\left( \rho _{B}^{\nu}\right) .
\end{align}%
\end{small}

Thus, taking $\delta \rightarrow 0$ for simplicity, we have
\begin{align}
\left\vert F_{\nu }\right\vert & =\left\vert T_{n}^{B,\nu }(P)\right\vert
/\left\vert M_{\nu }\right\vert {}  \notag \\
& =2^{n\left[ S\left( \Delta \rho _{B}^{\nu }\right) -I_{B:E}^{{}}\right] }{}
\notag \\
& =2^{n\left[ S\left( \Delta \rho _{B}^{\nu }\right) -S\left( \rho
_{B}^{\nu}\right) \right] }.
\end{align}

Let us relabel $\left( i^{n}\right) \rightarrow \left( f,m\right) $, then
the post-measurement state can be expressed as :
\begin{widetext}
\begin{align}
|\Phi _{ABE}^{\nu}\rangle _{T_{n}}^{\otimes n}& =\sqrt{P_{\nu }}|\Pi _{A}^{\nu
}\rangle ^{\otimes n}\frac{1}{\sqrt{\left\vert F_{\nu }\right\vert \cdot
\left\vert M_{\nu }\right\vert }}\sum_{f\in F_{\nu },m\in M_{\nu
}}\left\vert f\right\rangle \left\vert m\right\rangle \otimes \left\vert
\varphi _{E,v}^{fm}\right\rangle {}  \notag \\
& =\sqrt{P_{\nu}}|\Pi _{A}^{\nu }\rangle ^{\otimes n}\frac{1}{\sqrt{\left\vert
F_{\nu}\right\vert }}\sum_{f\in F_{\nu }}\left\vert f\right\rangle \frac{1}{%
\sqrt{\left\vert M_{\nu }\right\vert }}\sum_{m\in M_{\nu}}\left\vert
m\right\rangle \left\vert \varphi _{E,v}^{fm}\right\rangle ,
\end{align}%
\end{widetext}where $P_{\nu }=$Tr$\left( \Pi _{A}^{\nu }\otimes \mathbb{I}%
_{BC}\rho _{_{ABC}}\right) $. When we define $\left\vert \phi \right\rangle
_{\nu }^{f}\equiv \frac{1}{\sqrt{\left\vert M_{\nu }\right\vert }}\sum_{m\in
M_{\nu }}\left\vert m\right\rangle \left\vert \varphi
_{E,v}^{fm}\right\rangle $,~based on Uhlmann's theorem\thinspace \cite%
{uhlm1,uhlm2},~there exists a unitary $U_{\nu }^{f}$ on $E$ such that $%
\left( \mathbb{I}_{m}\otimes U_{\nu }^{f}\right) \left\vert \phi
\right\rangle _{\nu }^{f}\approx \left\vert \phi \right\rangle _{\nu }^{0}$
for each state $\left\vert \phi \right\rangle _{\nu }^{f}$. It implies that
we can construct the incoherence operation described by a group of Kraus
operators $\left\{ K_{r}^{\nu }\right\} $, satisfying $\sum_{r}{K_{r}^{\nu }}%
^{\dag }K_{r}^{\nu }=\mathbb{I}$, on each ensemble $\left\{ P_{\nu },|\Phi
_{ABE}^{\nu }\rangle _{T_{n}}^{\otimes n}\right\} \,$\cite{wintedistill},
i.e.,
\begin{equation}
K_{r}^{\nu }=\mathbb{I}_{A}^{\otimes n}\otimes \sum_{f\in F_{\nu
}}\left\vert f\right\rangle \left\langle f\right\vert \otimes \left\vert
0\right\rangle \left\langle r\right\vert U_{\nu }^{f}.
\end{equation}%
We obtain
\begin{equation}
K_{r}^{\nu }|\Phi _{ABE}^{\nu }\rangle _{T_{n}}^{\otimes n}\approx \sqrt{%
P_{\nu }}|\Pi _{A}^{\nu }\rangle ^{\otimes n}\frac{1}{\sqrt{\left\vert
F_{\nu }\right\vert }}\sum_{f\in F_{\nu }}\left\vert f\right\rangle \otimes
\left\vert 0\right\rangle \left\langle r|\phi _{\nu }^{0}\right\rangle .
\end{equation}%
Now we approximately obtain the maximal coherent state $\left\vert \Phi
_{B}^{\nu }\right\rangle _{\left\vert F_{\nu }\right\vert }=\frac{1}{\sqrt{%
\left\vert F_{\nu }\right\vert }}\sum_{f\in F_{\nu }}\left\vert
f\right\rangle $. With $n\rightarrow \infty $ and the majorization condition
$\Delta \left( \Phi _{B}^{\nu }\right) _{\left\vert F_{\nu }\right\vert
}\prec \Delta \left( \Phi _{2}^{\otimes nR}\right) \,$\cite{wintedistill},
where $\left( \Phi _{B}^{\nu }\right) _{\left\vert F_{\nu }\right\vert
}\equiv \left\vert \Phi _{B}^{\nu }\right\rangle _{\left\vert F_{\nu
}\right\vert }\left\langle \Phi _{B}^{\nu }\right\vert $, and $\Phi _{2}$
being the density of the 2-dimensional maximal coherent state,\ one has $%
\left( \Phi _{B}^{\nu }\right) _{\left\vert F_{\nu }\right\vert }\overset{%
\text{IO}}{\longmapsto }\Phi _{2}^{\otimes nR}$. There should be an equality
of the length of the typical sequences, i.e.,
\begin{equation}
\left\vert F_{\nu }\right\vert =2^{n\left[ S\left( \Delta \rho _{B}^{\nu
}\right) -S\left( \rho _{B}^{\nu }\right) \right] }=2^{nR_{\nu }S(\Delta
\Phi _{2})},
\end{equation}%
and thus%
\begin{equation}
R_{\nu }=\left[ S\left( \Delta \rho _{B}^{\nu }\right) -S\left( \rho
_{B}^{\nu }\right) \right] =C_{r}\left( \rho _{B}^{\nu }\right) ,
\end{equation}%
which means that with the assistance of the optimal coordinate measurement $%
\Pi _{A}^{\nu }$ we can distillate $\Phi _{2}$ by rate $R_{\nu }=C_{r}\left(
\rho _{B}^{\nu }\right) $ at the asymptotic limit. Finally, we have the
total distillation rate $\mathcal{R}=\sum_{\nu }P_{\nu }C_{r}\left( \rho
_{B}^{\nu }\right) =\underset{\{\Xi _{A}^{i}\}}{\max }\sum_{i}P_{i}C_{r}%
\left( \rho _{B}^{i}\right) .$ The proof is completed.

\subsection*{\textbf{APPENDIX C: PROOF OF THEOREM 2}}

Recalling the Theorem 2, i.e.,
\begin{equation}
C_{\text{CoP}}^{A|BC}\left( \Psi _{_{ABC}}\right) \geq C_{\text{CoP}%
}^{A|B}\left( \Psi _{_{ABC}}\right) +C_{\text{CoP}}^{A|C}\left( \Psi
_{_{ABC}}\right) ,
\end{equation}%
Let us give the detailed proof by defining the core: {\footnotesize $\tau \equiv C_{\text{%
CoP}}^{A|BC}\left( \Psi _{_{ABC}}\right) -C_{\text{CoP}}^{A|B}\left( \Psi
_{_{ABC}}\right) -C_{\text{CoP}}^{A|C}\left( \Psi _{_{ABC}}\right) {}$}.
For a general bipartite state, we have {\footnotesize $C_{\text{CoP}}^{A|B}(\rho _{_{AB}})\leq C_{d}^{A|B}(\rho _{AB})\leq
C_{r}^{A|B}(\rho _{AB})$}, and for the type of
states in Eq.\thinspace (\ref{tripartite-special}), we have $C_{\text{CoP}}^{A|BC}\left( \Psi
_{_{ABC}}\right) =S\left( \Delta \rho _{_{BC}}\right) $ , then there is an inequality:
\begin{equation}
\tau \geq S\left( \Delta \rho _{_{BC}}\right) -C_{r}^{A|B}\left( \rho
_{_{AB}}\right) -C_{r}^{A|C}\left( \rho _{_{AC}}\right) {}.  \label{tao-3-1}
\end{equation}%
It is known that relative entropy will increase by adding a
subsystem\thinspace \cite{rev.mod.entropy}, i.e., $C_{r}^{AB|C}(\Psi
_{ABC}){}{}\geq C_{r}^{A|C}\left( \rho _{_{AC}}\right) $, thus
\begin{equation}
\tau \geq S\left( \Delta \rho _{_{BC}}\right) -C_{r}^{A|B}\left( \rho
_{_{AB}}\right) -C_{r}^{AB|C}(\Psi _{ABC}){}{}  \label{tao-3-2}
\end{equation}%
By using the conditional entropy $S_{C|AB}$, we have $C_{r}^{A|B}\left( \rho
_{_{AB}}\right) +C_{r}^{AB|C}(\Psi _{ABC})=S_{C|AB}(\triangle ^{C}\Psi
_{ABC})+S(\triangle ^{B}\rho _{AB}).$ Since relative entropy cannot be
increased by performing CPTP operations, thus we have $S(\triangle ^{C}\rho
_{_{ABC}}\Vert \rho _{_{AB}}\otimes \triangle ^{C}\rho _{_{C}})\geq
S(\triangle ^{BC}\rho _{_{ABC}}\Vert \triangle ^{B}\rho _{_{AB}}\otimes
\triangle ^{C}\rho _{_{C}})$. By expanding the relative entropy,\ one will
obtain the relationship between the conditional entropy, i.e., \ $%
S_{C|AB}(\triangle ^{BC}\rho _{_{ABC}})\geq S_{C|AB}(\triangle ^{C}\rho
_{_{ABC}})$. Then \bigskip the right hand side (RHS) of the
inequality\thinspace (\ref{tao-3-2}) holds:
\begin{equation}
\text{RHS}\geq S\left( \Delta \rho _{_{BC}}\right) -S_{C|AB}(\triangle
^{BC}\Psi _{ABC})-S(\triangle ^{B}\rho _{AB})=0,  \label{tao-3-3}
\end{equation}%
which is due to
\begin{eqnarray}
&&S_{C|AB}(\triangle ^{BC}\Psi _{ABC})+S(\triangle ^{B}\rho _{AB})  \notag \\
&=&S(\triangle ^{BC}\Psi _{ABC})-S(\triangle ^{B}\rho _{AB})+S(\triangle
^{B}\rho _{AB})  \notag \\
&=&S(\triangle ^{BC}\Psi _{ABC})=S\left( \Delta \rho _{_{BC}}\right) .
\end{eqnarray}%
Finally, we have
\begin{equation}
\tau \geq 0.
\end{equation}%
Now let us extend the proof to the multipartite cases of $N>3$. For a pure
state $|\Psi \rangle _{AB_{1}...B_{N}}$, with the dimension of the auxiliary
is dim$(\mathcal{H}_{A})=2$, and its Schmidt decomposition\ can be presented
as:
\begin{align}
&|\Psi \rangle _{_{AB_{1}...B_{N}}}{}  \notag \\
&=\sqrt{\lambda _{1}}|\phi _{A}^{1}\rangle \sum\limits_{\{i_{B}^{\prime
}\}}|i_{B_{1}}^{\prime }...i_{B_{N}}^{\prime }\rangle +\sqrt{\lambda _{2}}%
|\phi _{A}^{2}\rangle |i_{B_{1}}...i_{B_{N}}\rangle ,
\end{align}%
where $\{i_{B}^{\prime }\}$ denotes the subset consisting of the reference basises
different from $|i_{B_{1}}...i_{B_{N}}\rangle $, and the two states of
auxiliary satisfy $\left\langle \phi _{A}^{2}|\phi _{A}^{1}\right\rangle =0$%
, then we have {\footnotesize $C_{\text{CoP}}^{A|B_{1}...B_{N}}(\Psi
_{AB_{1}...B_{N}})=S\left( \Delta ^{B_{1}...B_{N}}\Psi
_{AB_{1}...B_{N}}\right) $} with the definition $\Psi _{AB_{1}...B_{N}}\equiv
\left\vert \Psi _{AB_{1}...B_{N}}\right\rangle \left\langle \Psi
_{AB_{1}...B_{N}}\right\vert $.\ By using the tripartite
inequality\thinspace (\ref{tao-3-2}) and (\ref{tao-3-3}), we have
\begin{align}
&C_{\text{CoP}}^{A|B_{1}...B_{N}}(\Psi _{AB_{1}...B_{N}}){}  \notag \\
&\geq C_{r}^{A|B_{1}}(\rho _{_{AB_{1}}})+C_{r}^{AB_{1}|B_{2}...B_{N}}(\Psi
_{AB_{1}...B_{N}}),
\end{align}%
where {\footnotesize $C_{r}^{AB_{1}|B_{2}...B_{N}}(\Psi _{AB_{1}...B_{N}})=S\left( \Delta
^{B_{2}...B_{N}}\Psi _{AB_{1}...B_{N}}\right) $}. Then the tripartite
inequality is reused that
\begin{align}
&C_{r}^{AB_{1}|B_{2}...B_{N}}(\Psi _{AB_{1}...B_{N}}) {}  \notag \\
&\geq C_{r}^{AB_{1}|B_{2}}(\rho
_{_{AB_{1}B_{2}}})+C_{r}^{AB_{1}B_{2}|B_{3}...B_{N}}(\Psi _{AB_{1}...B_{N}})
\notag \\
&\geq C_{r}^{A|B_{2}}(\rho
_{_{AB_{1}B_{2}}})+C_{r}^{AB_{1}B_{2}|B_{3}...B_{N}}(\Psi _{AB_{1}...B_{N}}),
\end{align}%
where we use the property of the relative entropy $C_{r}^{AB_{1}|B_{2}}(\rho
_{_{AB_{1}B_{2}}})\geq C_{r}^{A|B_{2}}(\rho _{_{AB_{1}B_{2}}})$. Then

\begin{widetext}
\begin{eqnarray}
C_{r}^{AB_{1}B_{2}|B_{3}...B_{N}}(\Psi _{AB_{1}...B_{N}}) &\geq
&C_{r}^{AB_{1}B_{2}|B_{3}}(\rho
_{_{AB_{1}B_{3}}})+C_{r}^{AB_{1}B_{2}B_{3}|B_{4}...B_{N}}(\Psi
_{AB_{1}...B_{N}})  \notag \\
&\geq &C_{r}^{A|B_{3}}(\rho
_{_{AB_{1}B_{2}}})+C_{r}^{AB_{1}B_{2}B_{3}|B_{4}...B_{N}}(\Psi
_{AB_{1}...B_{N}}).
\end{eqnarray}%
\end{widetext}
By repeatedly using the inequalities above, one will finally have
\begin{align}
&C_{\text{CoP}}^{A|B_{1}...B_{N}}(\Psi _{AB_{1}...B_{N}}){}  \notag \\
&\geq \sum_{\alpha =1}^{N}C_{r}^{A|B_{\alpha }}\left( \rho
_{_{AB_{1}B_{2}\cdot \cdot \cdot B_{N}}}\right) {}  \notag \\
&\geq \sum_{\alpha =1}^{N}C_{\text{CoP}}^{A|B_{\alpha }}\left( \rho
_{_{AB_{1}B_{2}\cdot \cdot \cdot B_{N}}}\right) .
\end{align}%

Then the proof is completed.

\subsection*{ \textbf{APPENDIX D: PROOF OF THEOREM 3}}

For a multipartite state $\rho _{_{AB_{1}B_{2}\cdot \cdot \cdot B_{N}}}$, a
set of projective measurements $\{\Xi _{A}^{i}\}$ are performed on the
subsystem $A$ and classical communications are allowed among the subsystems,
then the residual states of each subsystem $B_{j}$ is $\rho _{_{B_{j}}}^{i}$%
. Assuming that an optimal set of operations $\{\tilde{\Xi}_{A}^{i}\}$ help
us to achieve the maximal average of the coherence, i.e., $\sum_{i}\tilde{P}%
_{i}\left[ \sum_{j=1}^{N}C_{r}\left( \tilde{\rho}_{_{B_{j}}}^{i}\right) %
\right] $. Because of the supper additivity of coherence relative entropy,
i.e., $C_{r}\left( \rho _{1}\right) +C_{r}\left( \rho _{2}\right) \leq
C_{r}\left( \rho _{12}\right) $, where the reduced density $\rho _{1\left(
2\right) }=\mathrm{Tr}_{2\left( 1\right) }\left( \rho _{12}\right) $. One
has
\begin{equation}
\sum_{i}\tilde{P}_{i}\left[ \sum_{j=1}^{N}C_{r}\left( \tilde{\rho}%
_{_{B_{j}}}^{i}\right) \right] \leq \sum_{i}\tilde{P}_{i}C_{r}\left( \tilde{%
\rho}_{_{B_{1}B_{2}\cdot \cdot \cdot B_{N}}}^{i}\right) .
\end{equation}%
Obviously, $\{\tilde{\Xi}_{A}^{i}\}$ is the optimal measurement to obtain
the maximum of the subsystem coherence $\sum_{j=1}^{N}C_{r}\left( \rho
_{_{B_{j}}}^{i}\right) $ and not the coherence of the composite system $%
C_{r}\left( \rho _{_{B_{1}B_{2}\cdot \cdot \cdot B_{N}}}^{i}\right) $, thus
when we take into account all the projective measurements $\{\Xi _{A}^{i}\}$%
, there is the following inequality, i.e.,
\begin{align}
\sum_{i}\tilde{P}_{i}C_{r}\left( \tilde{\rho}_{_{B_{1}B_{2}\cdot \cdot \cdot
B_{N}}}^{i}\right) &\leq \max_{\{\Xi _{A}^{i}\}}\sum_{i}P_{i}C_{r}\left( \rho
_{_{B_{1}B_{2}\cdot \cdot \cdot B_{N}}}^{i}\right){}\notag\\
 &=C_{\text{CoP}%
}^{A|B_{1}B_{2}\cdot \cdot \cdot B_{N}}(\rho _{_{AB_{1}B_{2}\cdot \cdot
\cdot B_{N}}}).\end{align}%
The proof is completed.
\subsection*{\textbf{APPENDIX E:} SELECTION OF OPTIMAL MEASUREMENT}
In this appendix, we show the details of how to choose the optimal
measurement performed on subsystem $A$. The cases of GHZ-type and $W$-type
states are considered. To obtain the assisted coherence distillation $C_{%
\text{CoP}}^{A|BC}$, the optimal measurement basises should be $\left(
|0\rangle \pm |1\rangle \right) /\sqrt{2}$, which is due to that both
GHZ-type and $W$-type states satisfy the Schmidt decomposition in
Eq.\thinspace (\ref{tripartite-special}). While, to obtain $C_{\text{CoP}%
}^{A|B}$ and $C_{\text{CoP}}^{A|C}$, one should take into account the
reduced density $\rho _{AB\text{ }}$and $\rho _{AC}$.

First, let us consider the case of $W$-type state. For the reduced
state $\rho _{AB}$, we perform a general projective measurement, with the
basis $|\varphi _{+}\rangle =\cos \theta |0\rangle +\sin \theta e^{i\varphi
}|1\rangle $ and $|\varphi _{-}\rangle =\sin \theta |0\rangle -\cos \theta
e^{i\varphi }|1\rangle $, on subsystem $A$.\ Then the corresponding
probability are $P_{+}=\left( 1+\cos ^{2}\theta \right) /3$ and $%
P_{-}=\left( 1+\sin ^{2}\theta \right) /3$, and the residual state of system
$B$ is (the classical communications between $A$ and $B$ are followed):
\begin{align}
\rho _{_{+,B}}& =\frac{3}{1+\cos ^{2}\theta }[\left( \frac{2\cos ^{2}\theta
}{3}p+\frac{\sin ^{2}\theta }{3}\right) \left\vert 0\right\rangle
\left\langle 0\right\vert {}  \notag \\
& +\frac{2}{3}(1-p)\left\vert 1\right\rangle \left\langle 1\right\vert {}
\notag \\
& +\sin \theta \cos \theta e^{-i\varphi }\frac{\sqrt{2}}{3}\sqrt{1-p}%
\left\vert 0\right\rangle \left\langle 1\right\vert {}  \notag \\
& +\sin \theta \cos \theta e^{i\varphi }\frac{\sqrt{2}}{3}\sqrt{1-p}%
\left\vert 1\right\rangle \left\langle 0\right\vert ].
\end{align}%
Then, we make use of $l_{1}$ norm (defined as $\mathcal{C}%
_{l_{1}}=\sum\nolimits_{i\neq j}\left\vert \rho _{i,j}\right\vert $, with $%
\rho _{i,j}$ being the off-diagonal elements) to measure the quantum
coherence. Through numerical calculation, we find that the behavior of the $%
l_{1}$ norm of coherence is similar with the relative entropy of coherence,
and the former is easy to calculate. For the residual state $\rho _{_{+,B}}$
and $\rho _{_{-,B}}$, we have that
\begin{equation}
\mathcal{C}_{l_{1}}(\rho _{_{+,B}})\sim \sqrt{1-p}\frac{\left\vert \sin
\theta \cos \theta \right\vert }{1+\cos ^{2}\theta },
\end{equation}%
and which is same with $\mathcal{C}_{l_{1}}(\rho _{_{-,B}})$. Obviously, the
average $l_{1}$ norm of coherence $\overline{C_{l_{1}}}=\sum%
\nolimits_{i=+,-}P_{i}\mathcal{C}_{l_{1}}(\rho _{_{i\text{,}B}})$ is
proportional to$\ \left\vert \sin \theta \cos \theta \right\vert $, which
means that $\overline{C_{l_{1}}}$ reaches its maximal value at $\theta =\pi
/4$, i.e., the optimal measurement basises on $A$ should be $\left(
\left\vert 0\right\rangle \pm \left\vert 1\right\rangle \right) /\sqrt{2}$.
The same is true for $\rho _{_{AC}}$.

To the case of the GHZ-type state, we first consider the reduced
state $\rho _{_{AC}}$ also by introducing a general form of the measurement
basises $|\varphi _{+}\rangle =\cos \theta |0\rangle +\sin \theta
e^{i\varphi }|1\rangle $ and $|\varphi _{-}\rangle =\sin \theta |0\rangle
-\cos \theta e^{i\varphi }|1\rangle $. It is easy to obtain that $%
P_{+}=P_{-}=1/2$. After measurement, the residual states of $C$ are $\rho
_{_{+,C}}$ and $\rho _{_{-,C}}$:
\begin{eqnarray}
\rho _{_{+,C}} &=&\left(
\begin{array}{cc}
1-p\sin ^{2}\theta & \sqrt{p(1-p)}\sin ^{2}\theta \\
\sqrt{p(1-p)}\sin ^{2}\theta & p\sin ^{2}\theta%
\end{array}%
\right) , \\
\rho _{_{-,C}} &=&\left(
\begin{array}{cc}
1-p\cos ^{2}\theta & p(1-p)\cos ^{2}\theta \\
\sqrt{p(1-p)}\cos ^{2}\theta & p\cos ^{2}\theta%
\end{array}%
\right) .
\end{eqnarray}%
Then the eigenvalues of the two density are $\left( 1\pm \sqrt{1-p^{2}\sin
^{2}2\theta }\right) /2$, and finally we have the average relative entropy
of coherence, i.e., the assisted distillable coherence of $\rho _{_{AC}}$:
\begin{equation}
2C_{\mathrm{CoP}}^{A|C}(\rho _{_{AC}}) =\max_{\theta }F(p,\theta ),
\end{equation}
where
\begin{widetext}
\begin{equation}
F(p,\theta ) =H\left\{ 1-p\sin ^{2}\theta ,\,p\sin ^{2}\theta \right\}
+H\left\{ 1-p\cos ^{2}\theta ,\,p\cos ^{2}\theta \right\} %
-2H\left\{ \frac{1}{2}\left( 1\pm \sqrt{1-p^{2}\sin ^{2}2\theta }\right)
\right\},
\end{equation}%
\end{widetext}
with {\footnotesize$H\left\{ A,B\right\} \equiv -(A\log A+B\log B)$} is the binary Shannon
entropy. By calculating the first and second order derivative of $F(p,\theta
)$ with respect to $\theta $, one can find that the minimum of $F(p,\theta )$
is at $\theta =\pi /4$, while the maximum can be reached at $\theta =0$ or $%
\pi $, which implies that the best measurement basises are $\left\{
|0\rangle \text{, }|1\rangle \right\} $, then $C_{\mathrm{CoP}}^{A|C}(\rho
_{_{AC}})=\frac{1}{2}H\{1-p,p\}.$

By doing a similar analysis to $\rho _{_{AB}}$, the maximal value of $C_{%
\mathrm{CoP}}^{A|B}(\rho _{_{AB}})$ can be reached at $\theta =\frac{\pi }{4}
$. Then the optimal measurement basises are $\left\{ \left( \left\vert
0\right\rangle +\left\vert 1\right\rangle \right) /\sqrt{2}\text{, }\left(
\left\vert 0\right\rangle -\left\vert 1\right\rangle \right) /\sqrt{2}%
\right\} .$\newline

\subsection*{APPENDIX F: MEASURES OF GENUINE TRIPARTITE QUANTUM CORRELATION $\Delta
_{\text{SEF}}$, AND $\mathcal{D}^{(3)}$}

In this appendix we introduce the concept of two types of genuine tripartite
quantum correlation. \bigskip The first one is based on the squared
entanglement of formation, i.e.,\thinspace \thinspace \cite%
{multi-formation-Bai}%
\begin{equation*}
\Delta _{\text{SEF}}\left( \rho _{_{ABC}}\right) =E_{f}^{2}(\rho
_{_{A|BC}})-E_{f}^{2}(\rho _{_{A|B}})-E_{f}^{2}(\rho _{_{A|C}}),
\end{equation*}%
which detects that the multipartite entanglement not stored in pairs of
qubits. $E_{f}(\rho _{_{i|j}})$ is the entanglement of formation in the
subsystem $\rho _{_{ij}}$ with the definition $E_{f}(\rho _{_{i|j}})=%
\underset{\left\{ p_{m},\left\vert \phi \right\rangle _{m}^{ij}\right\} }{%
\min }\sum_{m}p_{m}S\left( \mathrm{Tr}_{i}(\left\vert \phi \right\rangle
_{m}^{ij})\right) $, where the minimum is taken over all the pure state
decompositions $\left\{ p_{m},\left\vert \phi \right\rangle
_{m}^{ij}\right\} $. In two-qubit quantum states, the entanglement of
formation has an analytical expression $E_{f}(\rho _{_{i|j}})=H\left\{ \frac{%
1\pm \sqrt{1-C^{2}(\rho _{_{ij}})}}{2}\right\} ,$ where $H(x)=-x\log
x-(1-x)\log (1-x)$ is the binary entropy and $C\left( \rho _{_{ij}}\right) $
is the concurrence of $\rho _{_{ij}}$. Moreover, in a tripartite pure state $%
|\psi \rangle _{_{ABC}}$, we have the relation $E_{f}^{2}(\rho _{_{A|BC}})=$
$S^{2}(\rho _{_{A}})$ in which $E_{f}(A|BC)$ is the entanglement of
formation in the partition $A|BC\,$\cite{multi-formation-Bai} and $S(\rho )$
is the von Neumann entropy.

Another concept is the multipartite discord with the definition (for the
tripartite case)\thinspace \cite{genuine-discord}:
\begin{equation}
\mathcal{D}^{(3)}\left( \rho \right) :=\mathcal{D}\left( \rho \right) -%
\mathcal{D}^{(2)}{}\left( \rho \right) ,
\end{equation}%
where $\mathcal{D}^{(3)}\left( \rho \right) $ describes the genuine
tripartite quantum correlation. Genuine correlations should contain all the
contributions that cannot be accounted for considering any of the possible
subsystems. $\mathcal{D}\left( \rho \right) \equiv T\left( \rho \right) -%
\mathcal{J}\left( \rho \right) $ is called the total quantum discord with
the total information (or correlation information) $T\left( \rho \right)
\equiv S(\rho \Vert \rho _{_{i}}\otimes \rho _{_{j}}\otimes \rho _{_{k}})$,
and the total classical correlation $\mathcal{J}\left( \rho \right) \equiv
\underset{P\{i,j,k\}}{\max }\left[ S\left( \rho _{_{i}}\right) -S\left( \rho
_{_{i|j}}\right) +S\left( \rho _{k}\right) -S\left( \rho _{_{k|ij}}\right) %
\right] $ with the maximum among the 6 indices permutations of the
probability $P_{i,j,k}=P_{i|j,k}P_{j|k}P_{k}$. Note that $S\left( \rho
_{_{i|j}}\right) \equiv \underset{\{E_{l}^{i}\}}{\min }S\left(
i|\{E_{l}^{j}\}\right) $ with respect to the positive operator valued
measure (POVM) $\{E_{l}^{j}\}$, and the average entropy $S\left(
i|\{E_{l}^{j}\}\right) =\sum_{k}p_{k}S(\rho _{_{i|E_{m}^{j}}})$ with the
probability $p_{k}=\mathrm{Tr}(E_{l}^{j}\otimes \mathbb{I}\rho _{_{ij}})$
and the residual density $\rho _{_{i|E_{m}^{j}}}$. Extending to the
tripartite case, it becomes $S\left( \rho _{_{k|ij}}\right) \equiv \underset{%
\{E_{l}^{i},E_{l}^{j}\}}{\min }S\left( k|\{E_{l}^{i},E_{l}^{j}\}\right) $.
The minimum bipartite discord $\mathcal{D}^{(2)}{}\left( \rho \right) $ is
defined as $\mathcal{D}^{(2)}{}\left( \rho \right) \equiv \min [\mathcal{D}%
\left( \rho _{_{i,j}}\right) ,\mathcal{D}\left( \rho _{_{k,j}}\right) ,%
\mathcal{D}\left( \rho _{_{i,k}}\right) ]$. The symmetrized quantum discord$%
\ \mathcal{D}\left( \rho _{_{i,j}}\right) \equiv \min [\mathcal{D}\left(
\rho _{i:j}\right) ,\mathcal{D}\left( \rho _{j:i}\right) ]$, where $\mathcal{%
D}\left( \rho _{i:j}\right) \equiv I(\rho _{_{i,j}})-\underset{\{E_{m}^{j}\}}%
{\max }[S(\rho _{_{i}})-S(\rho _{_{i}}|\{E_{m}^{j}\})]$ is the quantum
discord, and $I(\rho _{_{i,j}})\equiv S(\rho _{_{i,j}}\Vert \rho
_{_{i}}\otimes \rho _{_{j}})$ is the mutual information. For the pure state $%
\left\vert \phi \right\rangle _{ijk}$, if the following inequality is
satisfies: $I(\rho _{_{ij}})\geq I(\rho _{_{ik}})\geq I(\rho _{_{jk}}),$
there is a simple result that $\mathcal{D}^{(3)}\left( \rho \right) =S(\rho
_{_{k}})$\thinspace \cite{genuine-discord}. Therefore, for the GHZ-type
states, it is easy to check by numerical calculation that $I(\rho
_{_{AB}})\geq I(\rho _{_{AC}})\geq I(\rho _{_{BC}})$. Then we have $\mathcal{%
D}_{\text{GHZ}}^{(3)}=S(\rho _{_{C}})=H\left\{ 1-\frac{p}{2},\frac{p}{2}%
\right\} $, and the minimum value $\mathrm{min}\mathcal{D}_{\text{GHZ}%
}^{(3)}=H\left\{ 1-\frac{p}{2},\frac{p}{2}\right\} \bigg|_{p=0}=0$, while
the maximum value $\mathrm{\max }\mathcal{D}_{\text{GHZ}}^{(3)}=H\left\{ 1-%
\frac{p}{2},\frac{p}{2}\right\} \bigg|_{p=1}=1$.

As for the assisted distillable coherence, one can analytically obtain the
minimum value of $\tau $ at $p=0$, where the distillable coherence $C_{%
\mathrm{CoP}}^{A|BC}(\rho _{_{ABC}})=1$, $C_{\mathrm{CoP}}^{A|C}(\rho
_{_{ABC}})=1$, and $C_{\mathrm{CoP}}^{A|B}(\rho _{_{ABC}})=0$, then $\tau =0$%
. While, at $p=1$, we have $C_{\mathrm{CoP}}^{A|BC}(\rho _{_{ABC}})=1$, $C_{%
\mathrm{CoP}}^{A|C}(\rho _{_{ABC}})=0$, and $C_{\mathrm{CoP}}^{A|B}(\rho
_{_{ABC}})=0$, which means $\tau =1$. In Fig.\thinspace 2(b), the numerical
and experimental results of $\tau $ show that in the case of the GHZ-type
state, the behaviors of $\tau $ and\ $\mathcal{D}_{\text{GHZ}}^{(3)}$ are
the same.

For the$\ W$-type states, the behavior of $\mathcal{D}^{(3)}$ is different
from the GHZ-type states that when $0\leq p\leq 0.5$, there are $I(\rho
_{_{AB}})\geq I(\rho _{_{BC}})\geq I(\rho _{_{AC}})$, then the tripartite
quantum discord $\mathcal{D}^{(3)}=S(\rho _{_{C}})$. When $0.5<p\leq 1$,
there are $I(\rho _{_{AC}})\geq I(\rho _{_{BC}})\geq I(\rho _{_{AB}})$, and
thus $\mathcal{D}^{(3)}=S(\rho _{_{B}})$. Obviously, on both sides of the
point $p=0.5$, $\mathcal{D}^{(3)}$ behaves differently. We also discuss the
relation between $\tau $ and $\mathcal{D}^{(3)}$. When $p=0$ and $1$, there
will be $\mathcal{D}^{(3)}=0$, where one can also find $\tau =0$. While, at
the special point of $p=0.5$, the two quantities reach their maximum values $%
\mathcal{D}^{(3)}\simeq 0.918$ and $\tau \simeq 0.848$. More clearly, one
can find the numerical and experimental results in Fig.\thinspace 2(a),
where $\tau $ and $\mathcal{D}^{(3)}$ display a similar behavior except the
regions near the maximal value.

\end{document}